\documentclass[prd,aps,floats,floatfix,superscriptaddress,preprintnumbers,
showpacs,eqsecnum,nofootinbib,twocolumn]{revtex4}
\usepackage{latexsym,array,theorem,mathrsfs,bm,float}

\usepackage{psfrag}
\usepackage{amsfonts,amsmath,amssymb,latexsym,array,afterpage,theorem,mathrsfs,bm,float,epsfig,color,graphicx,tabularx,here,multirow}
\newcommand{\nn}{\nonumber \\}
\newcommand{\bea}{\begin{eqnarray}}
\newcommand{\ena}{\end{eqnarray}}
\newcommand{\beann}{\begin{eqnarray*}}
\newcommand{\enann}{\end{eqnarray*}}

\newcommand{\gsim}{\, \mbox{\raisebox{-1.ex}
{$\stackrel{\textstyle>}{\textstyle\sim}$}}\,}
\newcommand{\lsim}{\, \mbox{\raisebox{-1.ex}
{$\stackrel{\textstyle<}{\textstyle\sim}$}}\,}
\newcommand{\ma}[1]{\mbox{$\mathcal{#1}$}}

\newcommand{\calhR}[1]{\raisebox{2ex}{\tiny ({\em h})}\hspace{-0.8em}{\ma R}}

\begin{document}

%<<<<<<<<<<<<< TITLE >>>>>>>>>>>>>>>%
\title{
Cosmological Dynamics of D-BIonic and DBI Scalar Field\\ and 
Coincidence Problem of Dark Energy
}

%<<<<<<<<<<<<< AUTHOR >>>>>>>>>>>>>>>%

\author{Sirachak {\sc Panpanich}}
\email{sirachakp-at-gravity.phys.waseda.ac.jp}
\address{Department of Physics, Waseda University, 
Okubo 3-4-1, Shinjuku, Tokyo 169-8555, Japan}
\author{Kei-ichi {\sc Maeda}}
\email{maeda-at-waseda.jp}
\address{Department of Physics, Waseda University, 
Okubo 3-4-1, Shinjuku, Tokyo 169-8555, Japan}
\author{Shuntaro {\sc Mizuno}}
\email{shuntaro.mizuno-at-aoni.waseda.jp}
\address{Waseda Institute for Advanced Study, Waseda University,
1-6-1 Nishi-Waseda, Shinjuku, Tokyo 169-8050, Japan}

%<<<<<<<<<<<<< ADDRESS >>>>>>>>>>>>>>>%

%<<<<<<<<<<<<< DATE >>>>>>>>>>>>>>>%
\date{\today}

%======================================%
%<<<<<<<<<<<<< ABSTRACT >>>>>>>>>>>>>>>%
%======================================%
%%%%%%%%%%%%%%%%%%%%%%%%%%%%%%%%%%%%%%%%%%%
%%%%%%%%%%%%%%%%%%%%%%%%%%%%%%%%%%%%%%%%%%%
%%%%%%%%%%%%%%%%%%%%%%%%%%%%%%%%%%%%%%%%%%%
%%%%%%%%%%%%%%%%%%%%%%%%%%%%%%%%%%%%%%%%%%%
\begin{abstract}
We study the cosmological dynamics of D-BIonic and DBI scalar field, 
which is coupled to matter fluid. 
%D-Bionic is one of the screening mechanisms which suppresses fifth force by first order derivative of scalar field. This mechanism is analogous to the Vainshtein mechanism, namely there exists a scale distance in which fifth force is suppressed comparing to Newtonian force. 
For the exponential potential and the exponential couplings, 
we find a new analytic scaling solution 
%yielded 
yielding  the accelerated expansion of the Universe. 
%\color{blue} and keeping the ratio between the energy density of matter fluid and that of the scalar field constant.
%\color{black} 
Since it
% gives a scaling solution and 
is shown to be an attractor
%stable 
for some range of the coupling parameters, the density parameter of matter fluid can be the observed value,
as in the coupled quintessence with a canonical scalar field.  
Contrary to the usual coupled quintessence, where the value of matter couple giving observed density parameter
is too large to satisfy observational constraint from CMB, we show that the D-BIonic theory can give similar solution
with much smaller value of matter coupling.
As a result, 
together with the fact that the D-BIonic theory has a screening mechanism,  
the D-BIonic theory 
%with the observational constraints 
can 
solve the so-called coincidence problem as well as 
the dark energy problem.
\end{abstract}
%%%%%%%%%%%%%%%%%%%%%%%%%%%%%%%%%%%%%%%%%%%
%%%%%%%%%%%%%%%%%%%%%%%%%%%%%%%%%%%%%%%%%%%
%%%%%%%%%%%%%%%%%%%%%%%%%%%%%%%%%%%%%%%%%%%
%%%%%%%%%%%%%%%%%%%%%%%%%%%%%%%%%%%%%%%%%%%
%<<<<<<<<<<<<< PACS NUMBERS >>>>>>>>>>>>>>>%
%\pacs{
%} 

%\pacs{04.25.dg, 04.50.Kd, 04.70.Dy} 

\maketitle

%======================================%
%<<<<<<<<<<<< SECTION I  >>>>>>>>>>>>>>%
%======================================%
%%%%%%%%%%%%%%%%%%%%%%%%%%%%%%%%%%%%%%%%%%%
%%%%%%%%%%%%%%%%%%%%%%%%%%%%%%%%%%%%%%%%%%%
%%%%%%%%%%%%%%%%%%%%%%%%%%%%%%%%%%%%%%%%%%%
%%%%%%%%%%%%%%%%%%%%%%%%%%%%%%%%%%%%%%%%%%%
\section{Introduction}
%%%%%%%%%%%%%%%%%%%%%%%%%%%%%%%%%%%%%%%%%%%%%%%%%%%
%%%%%%%%%%%%%%%%%%%%%%%%%%%%%%%%%%%%%%%%%%%%%%%%%%%
%%%%%%%%%%%%%%%%%%%%%%%%%%%%%%%%%%%%%%%%%%%%%%%%%%%
After the discovery of accelerated expansion of the Universe \cite{Riess,Perlmutter}, 
one of the attempts to explain this mysterious 
phenomenon is an introduction of a scalar field which is a dynamical field rolling down on a potential. 
The model is called a quintessence model \cite{Caldwell}. This can be a solution to the dark energy problem 
by adding a new degree of freedom to the Universe. In addition to the dark energy problem, 
%the 
another mystery is the so-called coincidence problem, which 
 is why the amounts of dark energy and 
%of 
matter fluid
(including cold dark matter) are in the same order of magnitude \cite{Ade}. This problem indicates that there might be 
some interaction between them. Thus, the idea gives a new model called a coupled quintessence model \cite{Amendola}.
 This model contains the original quintessence mechanism but
  also gives a new solution called a scaling solution. 
  This scaling solution also provides not only the accelerated expansion of the universe but also 
  the density parameter of matter fluid does not vanish. 
  Since this scaling solution is an attractor, it will be realised naturally at the late time.
  As a result, this is one of the possible ways to solve the coincidence problem at the same time 
 with the dark energy problem. 
 However, the scaling solution is difficult to be realised 
because it 
%may require 
requires a large coupling constant \cite{Amendola,DEbook}. 

In addition, there is another problem to introduce a universal scalar field.
Since a scalar field couples to matter fluid, this leads to a new interaction force 
 between them, the so-called fifth force \cite{STbook},
 which has not been detected until now \cite{constraints}. 
In order to preserve a scalar field model coupled to matter fluid, 
there must be some screening mechanism to hide a new force
 from the observations on the ground and solar-system  experiments. 
The screening mechanism means that the fifth force is 
  suppressed comparing to Newtonian force in a highly dense region or close to a massive source, 
  whereas it recovers in a low density region or far from a gravitating source. 
   Namely, we recover general relativity (GR) or Newtonian gravity 
    at short distance from a massive source or in a highly dense region
    just as in an astrophysical scale.

Three groups of  the screening mechanisms have so far been proposed. 
The first group is the screening by a scalar field $\phi$ or its effective potential, which 
 consists of the chameleon mechanism \cite{chameleon1,chameleon2,chameleon3}, 
 the symmetron mechanism \cite{symmetron1,symmetron2}, 
 and the dilaton (Damour-Polyakov) mechanism \cite{dilaton1,dilaton2}. 
In the chameleon mechanism, mass of a scalar field depends 
on matter density, then a scalar field gets a large mass in a high density region such as on the Earth.  
This leads to a short range interaction of the fifth force. 
While in the symmetron mechanism or in the dilaton mechanism,
the  coupling parameter between a scalar field 
 and matter fluid depends on the minimum of the effective potential. 
 In a high density region, for example symmetron mechanism, the symmetry has not broken. Then the minimum of the effective potential is at zero value. As a result, the coupling parameter is equal to zero. Herewith, 
the scalar field decouples from matter fluid in highly density region. 
 
 The second group is the screening by the first-derivative of a scalar field, $\partial \phi$,  or the kinetic term of a scalar field,
 which includes the D-BIonic screening \cite{Burrage}, the kinetic screening $P(X)$ \cite{kinetic1,kinetic2}, and 
 the k-Mouflage mechanism \cite{Babichev}. In this group,
  the screening mechanism works by domination of some non-linear term in the equation of motion of 
 the scalar field. 
 Since the equation of motion consists of not only the linear term, which leads to the inverse-square ($r^{-2}$) fifth force, but also 
the non-linear term, which leads to a different form of the force, 
there exists some typical distance below which the non-linear term
dominates, whereas at larger distance from the source, the linear term becomes dominant. 
The fifth force is then screened at short distance from the source. 
This is analogous to the Vainshtein mechanism. 

The last group is the screening by the second-derivative of a scalar field, $\partial \partial \phi$,  
 or the so-called Vainshtein mechanism \cite{Vainshtein}. This mechanism is found in many models, 
for example, the Galileon gravity \cite{galileon}, the Horndeski theory \cite{Horndeski, Deffayet:2011gz,Kobayashi:2011nu}, and also the massive gravity \cite{massive1,massive2}. 
In these models, the Vainshtein mechanism works
in the similar way as we mentioned, namely, there exists some typical distance called the Vainshtein radius,
below which the non-linear term
 is dominant. As a result, GR recovers at a short distance.

Since a screening mechanism is important when we have a scalar field, 
in order to explain the coincidence problem as well as the dark energy problem, 
we study cosmological behaviour of a coupled quintessence model, in which 
a screening mechanism works.
In this paper, we focus 
%into 
on the D-BIonic screening mechanism. 
This may have another advantage in realisation of a scaling solution because 
there exists non-canonical kinetic term which changes the dynamics of  the scalar field.
It is interesting whether we find a scaling solution which satisfies the observational 
 constraints and becomes an attractor or not.  
The D-BIonic 
%gravity 
theory can reduce to a coupled quintessence model under non-relativistic limit of the Lorentz factor
 (we will see clearly in the next section). This is the same as the DBI 
%gravity 
theory
  considered as a generalised quintessence model. 
 In the  DBI 
%gravity 
theory, we find the accelerating universe even though the Lorentz factor is much larger than unity \cite{Martin:2008xw,Copeland}
 that is why we call it generalised quintessence. 
Here we will analyse unifiedly both D-BIonic and  DBI 
%gravity 
theories because  those can be described 
in the similar forms.
 
In Sec. \ref{basiceq} we show the basic equations for this work. 
 In Sec. \ref{analytic} we find 
  analytic solutions corresponding to two solutions in a coupled scaling quintessence: 
One case such that the potential term dominates and the other case both potential and matter density terms do contribute in the dynamics. 
We show stability analysis of these solutions in Sec. \ref{stability}, and comparing to the observational data in Sec. \ref{observations}.
Finally, Sec. \ref{conclusions} is devoted to conclusions and remarks.

%%%%%%%%%%%%%%%%%%%%%%%%%%%%%%%%%%%%%%%%%%%
%%%%%%%%%%%%%%%%%%%%%%%%%%%%%%%%%%%%%%%%%%%
%%%%%%%%%%%%%%%%%%%%%%%%%%%%%%%%%%%%%%%%%%%
%======================================%
%<<<<<<<<<<<< SECTION II  >>>>>>>>>>>>>>%
%======================================%
%%%%%%%%%%%%%%%%%%%%%%%%%%%%%%%%%%%%%%%%%%%
%%%%%%%%%%%%%%%%%%%%%%%%%%%%%%%%%%%%%%%%%%%
%%%%%%%%%%%%%%%%%%%%%%%%%%%%%%%%%%%%%%%%%%%

\begin{widetext}
\vskip 1cm
%%%%%%%%%%%%%%%%%%%%%%%%%%%%%%%%%%%%%%%%%%%
%%%%%%%%%%%%%%%%%%%%%%%%%%%%%%%%%%%%%%%%%%%
%%%%%%%%%%%%%%%%%%%%%%%%%%%%%%%%%%%%%%%%%%%
\section{Basic equations in D-BIonic and DBI 
%gravity 
theories}
\label{basiceq}
%%%%%%%%%%%%%%%%%%%%%%%%%%%%%%%%%%%%%%%%%%%
%%%%%%%%%%%%%%%%%%%%%%%%%%%%%%%%%%%%%%%%%%%
%%%%%%%%%%%%%%%%%%%%%%%%%%%%%%%%%%%%%%%%%%%

%%%%%%%%%%%%%%%%%%%%%%%%%%%%%%%%%%%%%%%%%%%
%%%%%%%%%%%%%%%%%%%%%%%%%%%%%%%%%%%%%%%%%%%
\subsection{Field Equations in D-BIonic and DBI 
%gravity 
theories}
%%%%%%%%%%%%%%%%%%%%%%%%%%%%%%%%%%%%%%%%%%%
%%%%%%%%%%%%%%%%%%%%%%%%%%%%%%%%%%%%%%%%%%%

We consider the following action
\begin{align}
S = \int d^4 x \sqrt{-g} \left[\frac{1}{2}R - \frac{1}{f(\phi)} \sqrt{1+ f(\phi) (\partial \phi)^2} 
%\right. \nonumber \\ \left.
+ \frac{1}{f(\phi)} - V(\phi)\right] + \int d^4 x \mathcal{L}_m (A^2 (\phi) g_{\mu\nu}, \psi_m) \,, \label{action}
\end{align}
where a scalar field, $\phi$, couples conformally to matter fluid, $\psi_m$, with a conformal factor $A(\phi)$. 
 $f(\phi)$ and $V(\phi)$ are  an inverse D3-brane-like tension and a potential, respectively. 
We will use the units of $\kappa^2 =  8\pi G=1$.
 We use the word ``like" here because the DBI theory is in the Jordan frame in which the scalar field does not couple to matter. 
 Therefore, the action we are considering here is just an action contained non-canonical kinetic term or the DBI-like action.

Varying the action (\ref{action}) with respect to the metric and the scalar field, 
we obtain the field equations as follows:
\begin{align}
&
R_{\mu\nu} - \frac{1}{2} g_{\mu\nu} R = 
T^{\rm (m)}_{\mu\nu} + T^{(\phi)}_{\mu\nu}
\,, \label{fieldeq}
\\
&
\square \phi + \frac{f_{,\phi}}{f^2}\left[1 + f (\partial\phi)^2\right] - \frac{f_{,\phi}}{2f}(\partial\phi)^2 
%\nonumber \\ &
- \frac{1}{2[1 + f (\partial\phi)^2]}\left[
f_{,\phi}(\partial\phi)^4 + f \nabla_{\mu} (\partial\phi)^2 \nabla^{\mu} \phi\right]
\nonumber \\
&
~~~~
- \left(\frac{f_{,\phi}}{f^2} + V_{,\phi}\right)\sqrt{1 + f (\partial\phi)^2} 
%\nonumber \\ &
= - \frac{A_{,\phi}}{A} T^{\rm (m)} \sqrt{1 + f (\partial\phi)^2} \,,  \label{eom1}
\end{align}
where the symbol $_{,\phi} \equiv d/d\phi$ and $T^{\rm (m)}=T^{\rm (m)\mu}_{~~~\mu}$. The energy-momentum tensor of the scalar field is given by
\begin{equation}
T^{(\phi)}_{\mu\nu} \equiv  \frac{\partial_{\mu}\phi \partial_{\nu}\phi}{\sqrt{1+ f (\partial\phi)^2}} - g_{\mu\nu} 
\left[f^{-1}\sqrt{1+ f (\partial\phi)^2} - f^{-1} + V\right]
\,. \label{Tmunu}
\end{equation}
\end{widetext}
It gives the DBI 
%gravity 
theory for $f(\phi) > 0$, 
 while when $f(\phi) < 0$, it yields D-BIonic 
%gravity 
theory.

We assume that the conformal factor is given by the exponential form:
\begin{equation*}
A(\phi) = e^{g \phi} \,,
\end{equation*}
where $g$ is a coupling constant. 

According to the original D-BIonic 
%gravity 
theory \cite{Burrage},
 the inverse D3-brane-like tension is a negative constant, i.e., 
$f(\phi) = -\Lambda^{-4}$, where $\Lambda$ is a characteristic mass scale, 
thus $f_{,\phi} =0$. The equation for the scalar field is simplified as
\begin{equation}
\nabla_{\mu} \left(\frac{\nabla^{\mu}\phi}{\sqrt{1 -\Lambda^{-4} (\partial\phi)^2}}\right)- V_{,\phi}= -  g T^{\rm (m)}.
\label{eom2}
\end{equation}
This is the same equation of motion as Eq.~(\ref{eom0}) except the potential term. The potential is necessary for studying cosmology as we will see in Sec. \ref{analytic}

Eq.~(\ref{eom2}) obviously consists of a linear term and a non-linear term, then there must be a characteristic radius 
 analogous to the Vainshtein radius.
Below this radius, we find a screening mechanism, which is called D-BIonic screening  \cite{Burrage}.

%%%%%%%%%%%%%%%%%%%%%%%%%%%%%%%%%%%%%%%%%%%%%
%%%%%%%%%%%%%%%%%%%%%%%%%%%%%%%%%%%%%%%%%%%%%
\subsection{Basic Equations for Coupled D-BIonic and DBI Cosmology}
\label{basic_equations}
%%%%%%%%%%%%%%%%%%%%%%%%%%%%%%%%%%%%%%%%%%%%%
%%%%%%%%%%%%%%%%%%%%%%%%%%%%%%%%%%%%%%%%%%%%%
In order to study the evolution of the Universe, 
we assume that the scalar field is homogeneous, namely $\phi =\phi(t)$
 and the spacetime is 
described by the flat Friedmann-Lemaitre-Robertson-Walker (FLRW) metric:
\begin{equation*}
ds^2 = - dt^2 + a^2 (t) d \bf x^2 \,. \nonumber
\end{equation*}
Consequently, 
%the 
Eq.~(\ref{eom1}) becomes
\begin{equation}
\ddot \phi + \frac{3 H \dot \phi}{\gamma^2} + \frac{V_{,\phi}}{\gamma^3} + \frac{f_{,\phi}}{f} 
\frac{(\gamma+2)(\gamma-1)}{2\gamma(\gamma+1)} \dot \phi^2 = \frac{g}{\gamma^3} T^{\rm (m)} \,, 
\label{eom3}
\end{equation}
where we introduce 
the ``Lorentz factor'' as
\begin{equation}
\gamma \equiv \frac{1}{\sqrt{1 - f(\phi) \dot \phi^2}} \,. 
\label{lorentz}
\end{equation}
In the standard DBI theory ($f(\phi) > 0$), $\gamma$ takes the values from $1$ to $\infty$, 
while in the D-BIonic 
%model 
theory ($f(\phi) < 0$), 
$\gamma$ is limited in the range of $(0, 1)$ instead. 
From Eq.~(\ref{eom3}), in the limit of $\gamma = 1 $ $ (|f(\phi)| \dot \phi^2 \ll 1)$, 
it  obviously becomes the equation of motion for the coupled quintessence model. 
We then find the both limits of the Lorentz factor as
\begin{eqnarray*}
&&
\gamma_{\text{DBI}} =
\begin{cases}
\infty & : \text{when} \, f(\phi) \dot \phi^2 \simeq 1 \, \text{(relativistic limit)} \\
1 & : \text{the coupled quintessence}
\end{cases}
\\
&&
\gamma_{\text{D-BIonic}} =
\begin{cases}
1 & : \text{the coupled quintessence} \\
0 & : \text{when} \, -f(\phi) \dot \phi^2 \gg 1  \, 
\\
& ~~~\text{(``anti''-relativistic limit)} \\
\end{cases}
\end{eqnarray*}
Therefore, the DBI-like action (\ref{action}) is  generalisation of 
%a 
the coupled quintessence model.

From Eq.~(\ref{Tmunu}), the pressure and the energy density of the scalar field are
given by 
\begin{align*}
\rho_{\phi} &= \frac{\gamma^2}{\gamma+1} \dot \phi^2 + V(\phi) \,, \\ 
P_{\phi} &=  \frac{\gamma}{\gamma+1} \dot \phi^2 - V(\phi) \,.
\end{align*}
Subsequently, the Friedmann equation is given by 
\begin{align}
H^2 = \frac{1}{3}\left( \frac{\gamma^2}{\gamma+1}\dot \phi^2 + V(\phi)+ \rho_{\rm m}\right) \,, 
\label{Friedmann}
\end{align}
where $H=\dot a/a$.
$\rho_{\rm m}$ is the total matter density
 (non-relativistic matter $+$ radiation), which we
  combine just for simplicity in our description.

Since the scalar field couples to matter fluid, this leads to modification on the energy equation. 
Namely, neither the scalar field energy nor matter fluid energy is conserved 
(however the total energy is of course conserved). 
For conformally coupled case, we obtain
\begin{equation*}
\nabla_{\mu} T^{\rm (m) \,\mu}_{~~~~\nu} = \frac{A_{,\phi}}{A} T^{\rm (m)} \nabla_{\nu} \phi \,.
\end{equation*}
According to the equation of state (EOS) for the matter fluid, $P_{\rm m} = w \rho_{\rm m}$, 
the energy equation of matter density becomes
\begin{align}
%\dot \rho_{\phi} + 3H (\rho_{\phi} + P_{\phi}) &= -  g (1-3w) \rho_M \dot \phi, \\
\dot \rho_{\rm m} + 3H (1+ w) \rho_{\rm m} &=  g (1 - 3w) \rho_{\rm m} \dot \phi \,, 
\label{matter}
\end{align}
where $w$ is the EOS parameter of matter fluid ($w = 0$ for non-relativistic matter, and $w = 1/3$ for radiation).

The basic equations we will use many times in this work are the equation of motion (\ref{eom3}), 
the Friedmann equation (\ref{Friedmann}), and the energy equation of matter density (\ref{matter}).

It is worth mentioning here that for radiation the energy 
%equation 
is conserved because 
the electromagnetic field is conformally invariant. Thus, radiation still decreases with the rate $a^{-4}$ as the Universe expands. 
Because the coupling constant, $g$, acts on only non-relativistic matter, at late time, we
 ignore the radiation component in the Universe.

In the next section we will find analytic solutions of the D-BIonic and DBI 
%gravity 
theories.

%%%%%%%%%%%%%%%%%%%%%%%%%%%%%%%%%%%%%%%%%%%
%%%%%%%%%%%%%%%%%%%%%%%%%%%%%%%%%%%%%%%%%%%
%%%%%%%%%%%%%%%%%%%%%%%%%%%%%%%%%%%%%%%%%%%
%======================================%
%<<<<<<<<<<<< SECTION III  >>>>>>>>>>>>>>%
%======================================%
%%%%%%%%%%%%%%%%%%%%%%%%%%%%%%%%%%%%%%%%%%%
%%%%%%%%%%%%%%%%%%%%%%%%%%%%%%%%%%%%%%%%%%%

%%%%%%%%%%%%%%%%%%%%%%%%%%%%%%%%%%%%%%%%%%%
%%%%%%%%%%%%%%%%%%%%%%%%%%%%%%%%%%%%%%%%%%%
%%%%%%%%%%%%%%%%%%%%%%%%%%%%%%%%%%%%%%%%%%%
\section{ANALYTIC SOLUTIONS}
\label{analytic}
%%%%%%%%%%%%%%%%%%%%%%%%%%%%%%%%%%%%%%%%%%%
%%%%%%%%%%%%%%%%%%%%%%%%%%%%%%%%%%%%%%%%%%%
%%%%%%%%%%%%%%%%%%%%%%%%%%%%%%%%%%%%%%%%%%%

Here we shall discuss some particular forms of $f(\phi)$ and $V(\phi)$, i.e.,
 \begin{equation*}
f(\phi)=\epsilon f_0 e^{-\mu\phi} ~~{\rm and}~~~
V(\phi)=V_0 e^{-\lambda\phi}\,.
\end{equation*}
We assume $ f_0>0, V_0>0$.  We also assume $\lambda>0$ without loss of generality 
(If $\lambda<0$, redefining the scalar field as $\phi \rightarrow -\phi$, we find 
our action.).
The parameter $\epsilon=1$ gives the DBI 
%gravity 
theory, while  $\epsilon=-1$ 
%is 
gives the D-BIonic theory.

Since we are interested in the special form of the kinetic term in the D-BIonic or DBI theory,
we look for (asymptotic) solution with $\gamma = \gamma_0$ = constant.
This condition leads to $f(\phi)\dot\phi^2$ = constant.
From our ansatz of the function $f(\phi)$, we solve for (asymptotic) solution of 
the scalar field as 
\begin{equation*}
\phi=-{2\over \mu}\ln \left({t\over t_0}\right) +\phi_0
\,,
\end{equation*}
where $\phi_0$ is the value of $\phi$ at $t=t_0$. Taking derivatives with respect to time, we find
\begin{equation*}
\dot \phi = - \frac{2}{\mu t}. 
\end{equation*}
Clearly, $\dot \phi > 0$ when $\mu < 0$. This corresponds to the scalar field motion 
rolling down the runaway exponential potential. I

We assume that a scale factor increases as a power-law expansion:
\begin{equation*}
a=a_0\left({t\over t_0}\right)^p
\,.
\end{equation*}
This is natural because the kinetic term is proportional to $t^{-2}$.
If we do not assume a power-law expansion, the kinetic term does not play any important role in the spacetime dynamics, 
and then it gives the same results as those with the conventional canonical kinetic term.

\begin{widetext}
Consequently, assuming that matter is given only by dust fluid ($w=0$), 
the basic equations given in Sec. \ref{basic_equations} are reduced to be
\begin{align*}
&{2\over \mu t^2}-{6p\over \gamma_0^2\mu t^2}
-{\lambda V_0\over \gamma_0^3}e^{-\lambda \phi_0}
\left({t\over t_0}\right) ^{2\lambda\over \mu} 
%\nonumber \\
%& \quad
 -{2(\gamma_0+2)(\gamma_0-1)\over \mu \gamma_0(\gamma_0+1) t^2} = -{g\over \gamma_0^3}\rho_{\rm m}
\,, 
\\
%\end{align}
%\begin{align}
&{p^2\over t^2}={1\over 3}\left[
{4\gamma_0^2\over (\gamma_0+1)\mu^2 t^2}+V_0 e^{-\lambda
\phi_0}\left({t\over t_0}\right) ^{2\lambda\over \mu}+\rho_{\rm m}
\right]\,, \\
& \dot \rho_{\rm m} +{3 p\over t}\rho_{\rm m}=-{2g\over t}\rho_{\rm m}
\,.
\end{align*}
The last equation is easily integrated as
\begin{equation*}
\rho_{\rm m}=\rho_0\left({t\over t_0}\right)^{-q}
\,,
\end{equation*}
where we define
\begin{equation*}
q\equiv 3p+{2g\over \mu}
\,.
\end{equation*}
We  then rewrite the basic equations as
\begin{align}
V_0e^{-\lambda\phi_0}\left({t\over t_0}\right)^{2\lambda\over \mu}
&={-2\mu\gamma_0[3p(\gamma_0+1)-2\gamma_0]+g(3p^2(\gamma_0+1)\mu^2-4\gamma_0^2)\over \mu^2t^2(\gamma_0+1)(\lambda+g)} \,,
\label{Veom}
\\
\rho_0\left({t\over t_0}\right)^{-q}
&={2\mu\gamma_0[3p(\gamma_0+1)-2\gamma_0]+\lambda(3p^2(\gamma_0+1)\mu^2-4\gamma_0^2)\over \mu^2t^2(\gamma_0+1)(\lambda+g)} \,. 
\label{rhoeom}
\end{align}
\end{widetext}
Obviously, if either the term with $V_0$ or another one with $\rho_0$ is dominant, we do not find 
any asymptotic solution with our ansatz. In fact, if $\lambda /  \mu > -1$ or $q<2$, at late time we obtain 
$V_0= 0$ or $\rho_0=0$, which does not give any interesting solutions. 
Hence we consider  the cases with
\begin{equation*}
{\lambda\over \mu} \leq -1~~{\rm and}~~~q\geq 2
\,.
\end{equation*}

For the case with $\lambda /  \mu < -1$ or $q>2$, the potential term or matter term 
does not contribute asymptotically in the dynamics.
Then we shall classify the (asymptotic)  solutions into the following four cases: \\
{\bf I}. Both the potential term and the matter density do contribute in the dynamics ($\lambda/\mu=-1$ and $q=2$),
\\
{\bf II}. The matter density does not contribute, but the potential term does ($\lambda/\mu=-1$ and $q>2$), 
\\
{\bf III}. The potential term does not contribute, but the matter density does ($\lambda/\mu<-1$ and $q=2$), 
\\
{\bf IV}. Both the potential term and the matter density do not contribute in the dynamics ($\lambda/\mu<-1$ and $q>2$).

In this text, we discuss only the case that the potential plays an important role, i.e., the cases I and II.
These correspond to the scaling solution and the conventional quintessence solution in 
the coupled quintessence model, respectively.
In Appendix, we shall discuss the other two cases ({\bf III} and {\bf IV}).

%%%%%%%%%%%%%%%%%%%%%%%%%%%%%%%%%%%%%%%%%%%%%%%%%%%%%%%%
%%%%%%%%%%%%%%%%%%%%%%%%%%%%%%%%%%%%%%%%%%%%%%%%%%%%%%%%
\subsection{Case I : $\mu=-\lambda$ and $q=2$}
%%%%%%%%%%%%%%%%%%%%%%%%%%%%%%%%%%%%%%%%%%%%%%%%%%%%%%%%
%%%%%%%%%%%%%%%%%%%%%%%%%%%%%%%%%%%%%%%%%%%%%%%%%%%%%%%%
From the definition of $q$, we have 
\begin{equation*}
p={2\over 3}\left(1+{g\over \lambda}\right)
\,,
\end{equation*}
and using the equation of $p$, the full equations of (\ref{Veom}) and (\ref{rhoeom}) solve as
\begin{align}
V_0 e^{-\lambda\phi_0} t_0^2 &={4\gamma_0
\over \lambda^2(\gamma_0+1)}-2p+3p^2  \,,  
\label{VD1} \\
\rho_0t_0^2&=
-{4\gamma_0\over \lambda^2}+2p\,.  
\label{rhoD1}
\end{align}

Since $\gamma_0$ contains $e^{-\lambda\phi_0} t_0^2$
as 
\begin{equation*}
\gamma_0={1\over \sqrt{1-{4\epsilon f_0\over 
%\mu
\lambda^2e^{-\lambda\phi_0} t_0^2}}}
\,,
\end{equation*}
substituting $e^{-\lambda\phi_0} t_0^2$ into the Eq.~(\ref{VD1}),
 we obtain the equation for $\gamma_0$:
\begin{align}
&[3(1-\epsilon f_0V_0)
+g(g+\lambda)]\gamma_0^2
\nn
&~~~~~~~~~~
-3\gamma_0
-g(g+\lambda)=0
\,,
\label{gamma_I}
\end{align}
whose solution is given by 
\begin{align*}
\gamma_0 =\gamma_0^{(\pm)} \equiv {3\pm \sqrt{D} \over 2[3(1-\epsilon f_0V_0)
+g(g+\lambda)]}
\,,
\end{align*}
where the discriminant $D$ is defined by
\begin{equation*}
D= 4g(g+\lambda)[3(1 - \epsilon f_0V_0) + g(g+\lambda)] + 9
\,.
\end{equation*}
$D$ must be non-negative in order to find a real solution for $\gamma_0$.
For the D-BIonic ($\epsilon=-1$), it is always positive definite, then the root for $\gamma_0$ 
exists.
For the DBI ($\epsilon=+1$), we find the condition as
\begin{equation}
f_0V_0 \leq 1 + \frac{1}{3} g (g+\lambda) + \frac{3}{4g(g+\lambda)}
\,.~~~~
\label{con_DBI1}
\end{equation}

Since it turns out that $\gamma_0^{(-)}$  branch solution does not give the
accelerating Universe 
%always becomes
%negative 
for both DBI and D-BIonic, 
we then 
%have 
consider only $\gamma_0^{(+)}$ solution. 
For the DBI case, we obtain the additional condition otherwise even $\gamma_0^{(+)}$ gives 
%negative value
the value smaller than 1
\begin{equation}
f_0V_0 < 1 + \frac{1}{3} g (g+\lambda) \,,
\label{con_DBI2}
\end{equation} 
This condition is tighter than {the previous one (\ref{con_DBI1}), 
then the condition (\ref{con_DBI2}) always gives the non-negative discriminant.

From Eq. (\ref{rhoD1}) and the equation of $p$, we find 
\begin{align*}
\rho_0t_0^2&=
{4\left[
\lambda\left(g+\lambda \right)-3\gamma_0 \right]\over 3 
%\mu
\lambda^2}
\,, 
\end{align*}
this leads to the additional condition
\begin{equation}
\lambda\left(g+\lambda \right)-3 \gamma_0 \geq 0
\label{con_density}
\,.
\end{equation}
The above condition gives the constraint on the coupling constant for the existence of the solution 
as 
\begin{equation}
g\geq g_{\rm cr}
\,,
\label{con_A1}
\end{equation}
with
\begin{equation*}
g_{\rm cr}\equiv {\lambda\over 2}\left[-1+
\sqrt{\left(1- {6\over \lambda^2} \right)^2+{12\epsilon f_0V_0\over \lambda^2}}\right]
\,. 
\end{equation*}

In order to find an accelerating Universe, 
since 
\begin{equation*}
p=\frac{2}{3}\left(1 + \frac{g}{\lambda}\right) >1\,, \nonumber
\end{equation*}
we obtain
\begin{equation}
g>{\lambda\over 2} \,. 
\label{con_A2}
\end{equation}
Therefore, Eqs.~(\ref{con_A1}) and (\ref{con_A2}) are the conditions of $g$ for realising the scaling solution {\bf I}
giving an accelerating Universe. 
Eq.~(\ref{con_A1}) gives the tighter condition for $\lambda < \lambda_{\rm cr}$,
where   $\lambda_{\rm cr}$ is given by
\begin{equation*}
\lambda_{\rm cr}^2\equiv 
2\left[-(1-\epsilon f_0V_0)+\sqrt{(1-\epsilon f_0V_0)^2+3}\right]
\,,
\end{equation*}
while Eq.~(\ref{con_A2}) gives the tigher condition for  $\lambda > \lambda_{\rm cr}$,

The EOS parameter of the scalar field is given by
\begin{equation*}
w_\phi=-1+{3\gamma_0\over 3\gamma_0+g(g+\lambda)}
\,.
\end{equation*}
We also introduce the effective EOS parameter $w_{\rm eff}$ by
\begin{eqnarray*}
 w_{\rm eff}\equiv -1-{2\dot H\over 3H^2}=-1+{2\over 3p}.
\end{eqnarray*}
The present solution gives
\begin{eqnarray*}
 w_{\rm eff}=-{g\over g+\lambda}.
\end{eqnarray*}

The matter density and the scalar field density are scaled in this solution.
Then we can evaluate the asymptotic values of $\Omega_m$ and $\Omega_\phi$ as follows:
\begin{align*}
\Omega_{\rm m}&=
{\lambda(g+\lambda)-3\gamma_0\over (g+\lambda)^2} \,,
\\
\Omega_\phi&={g(g+\lambda)+3\gamma_0\over (g+\lambda)^2}
\,.
\end{align*}

We find the scaling solution {\bf I} for accelerating Universe by contributions from both potential and matter density, 
which is given by $\gamma_0^{(+)}$ and $p$, with the constraints on $g$. Note that there is another constraint (\ref{con_DBI2}) 
on $f_0 V_0$ for the DBI theory.

%%%%%%%%%%%%%%%%%%%%%%%%%%%%%%%%%%%%%%%%%%%%%%%%%%%%%%%%
%%%%%%%%%%%%%%%%%%%%%%%%%%%%%%%%%%%%%%%%%%%%%%%%%%%%%%%%
\subsection{Case II : $\mu=-\lambda$ and $q>2$}
%%%%%%%%%%%%%%%%%%%%%%%%%%%%%%%%%%%%%%%%%%%%%%%%%%%%%%%%
%%%%%%%%%%%%%%%%%%%%%%%%%%%%%%%%%%%%%%%%%%%%%%%%%%%%%%%%
In this case, the matter density does not contribute the dynamics asymptotically, 
the basic equations for the asymptotic solution (\ref{Veom}) and (\ref{rhoeom}) 
give Eq.~(\ref{VD1}) and 
\beann
-2\lambda\gamma_0[3p(\gamma_0+1)-2\gamma_0]
+\lambda(3p^2(\gamma_0+1)\lambda^2-4\gamma_0^2)=0
\,,
\enann
which gives
\begin{equation}
p={2\gamma_0\over \lambda^2}
\,, \label{pC2}
\end{equation}
unless $p=0$. 
Then we obtain from Eq.~(\ref{VD1})
\begin{equation}
V_0 e^{-\lambda \phi_0}t_0^2={4\gamma_0^2\over \lambda^4}\left[3-{\lambda^2\over \gamma_0+1}\right]
\,. \label{VC2}
\end{equation}
Since we assume $V_0>0$, we have a constraint 
\begin{equation*}
\lambda^2<3(\gamma_0+1)
\,.
\end{equation*}

Using the definition of $\gamma_0$, we eliminate $e^{-\lambda \phi_0}t_0^2$ in Eq.~(\ref{VC2}), 
and we find the equation for $\gamma_0$ as
\begin{equation}
3\gamma_0^2-\lambda^2\gamma_0+\lambda^2(1-\epsilon f_0V_0)-3=0 
\,. \label{gammaC2}
\end{equation}
Then $\gamma_0$ is given by 
\begin{align*}
&\gamma_0=\gamma_0^{(\pm)}
\nn
&~~\equiv
{\lambda^2\over 6}\left[1\pm \sqrt{1-2(1-\epsilon f_0V_0) \left({6\over \lambda^2}\right)
+\left({6\over \lambda^2}\right)^2}\right]
\,.
\end{align*}

The existence of the real roots for this equation, we find the condition such that 
\begin{equation*}
\left({\lambda^2\over 6}\right)^2-2(1-\epsilon f_0V_0) \left({\lambda^2\over 6}\right)
+1\geq 0
\,.
\end{equation*}
For the DBI 
%gravity 
($\epsilon = +1$), this condition is always satisfied, and then 
we can find the solution.
On the other hand, for the D-BIonic 
%gravity 
($\epsilon = -1$), 
we have the condition on $\lambda$ for the existence of the root.
We find 
\begin{equation*}
\lambda^2\geq \lambda_+^2~~{\rm or}~~~\lambda^2\leq \lambda_-^2\,,
\end{equation*}
where
\begin{equation*}
\lambda_\pm^2
\equiv 6\left[1-\epsilon f_0V_0\pm \sqrt{(1-\epsilon f_0V_0)^2-1}
\right]\,.
\end{equation*}
Our ansatz $q>2$ gives another constraint such that 
\begin{eqnarray*}
3\gamma_0>\lambda(g+\lambda)
\,,
\end{eqnarray*}
which is reduced to 
\begin{eqnarray*}
g&<&g_{\rm cr}
\,.
\end{eqnarray*}

For the power of expansion, $p$, substituting $\gamma_0$ into Eq. (\ref{pC2}), we find 
\begin{align*}
&p=p^{(\pm)}
\nn
&~~
\equiv
{1 \over 3}
\left[1\pm \sqrt{1-2(1-\epsilon f_0V_0) \left({6\over \lambda^2}\right)
+\left({6\over \lambda^2}\right)^2}\right]
\,.
\end{align*}
However, in order to obtain the accelerating Universe ($p>1$), only positive-branch ($\gamma_0^{(+)}$ and $p^{(+)}$)
is possible. We then find the condition for $p>1$ is given by
\begin{equation*}
3\left({\lambda^2\over 6}\right)^2+
2(1-\epsilon f_0V_0) \left({\lambda^2\over 6}\right)
-1<0\,,
\end{equation*}
i.e., 
\begin{equation*}
0<\lambda^2<\lambda_{\rm cr}^2\,,
\end{equation*}
where the critical value
$\lambda_{\rm cr}$ is the same as the one defined in the previous subsection.
%defined by
%\begin{equation*}
%\lambda_{\rm cr}^2\equiv 
%2\left[-(1-\epsilon f_0V_0)+\sqrt{(1-\epsilon f_0V_0)^2+3}\right]
%\,.
%\end{equation*}
This condition always satisfies the constraint of $\lambda^2\leq \lambda^2_{-}$ for the D-BIonic. Therefore, 
we have the accelerating Universe solution {\bf II} with $\gamma_0^{(+)}$ and $p^{(+)}$, 
where there is the upper bound $\lambda_{\rm cr}^2$.

Note that when $\epsilon f_0 V_0 =0$, we recover the conventional acceleration condition in the quintessence model such that 
$\lambda_{\rm cr}=\sqrt{2}$.  For the DBI theory, the constraint becomes weaker
($\lambda_{\rm cr}>\sqrt{2}$), while 
for the D-BIonic theory, it becomes stronger ($\lambda_{\rm cr}<\sqrt{2}$).

The EOS parameter of the scalar field in this case is given by
\begin{equation*}
%1+w^{(\phi)}={\gamma_0^2-1\over \gamma_0[(\gamma_0-1)+\epsilon f_0 V_0]}
w_\phi=-1+{\lambda^2\over 3\gamma_0}\,.
\end{equation*}
When $\gamma_0 = 1$, the $w_\phi$ is the same as that in the quintessence model 
\cite{Copeland:1997et}.
%(cite).

\begin{widetext}
%%%%%%%%%%%%%%%%%%%%%%%%%%%%%%%%%%%%%%%%%%%%%%%%%%%%%%%%
%%%%%%%%%%%%%%%%%%%%%%%%%%%%%%%%%%%%%%%%%%%%%%%%%%%%%%%%
\subsection{The solution {\bf I} or the solution {\bf II}}
\label{expectation}
%%%%%%%%%%%%%%%%%%%%%%%%%%%%%%%%%%%%%%%%%%%%%%%%%%%%%%%%
%%%%%%%%%%%%%%%%%%%%%%%%%%%%%%%%%%%%%%%%%%%%%%%%%%%%%%%%
Here first we summarise the above results in Table \ref{Table I} and Fig \ref{F1}.
\begin{table}[H]
\begin{center}
  \begin{tabular}{|c||c|c|c|c|}
\hline
&\multicolumn{2}{|c|}{} &  \multicolumn{2}{|c|}{}\\[-.5em]
 &\multicolumn{2}{|c|}{The solution {\bf I}} &  \multicolumn{2}{|c|}{The solution {\bf II}}\\[.2em]
\hline 
&& &&\\[-.5em]
Theory&  DBI ($\epsilon=1$)  &  ~~~~D-BIonic ($\epsilon=-1$)  ~~~~ & DBI ($\epsilon=1$)   &  D-BIonic  ($\epsilon=-1$)   
\\[.2em]
\hline \hline 
&\multicolumn{2}{|c|}{} &  \multicolumn{2}{|c|}{}\\[-.5em]
$\gamma_0$
&\multicolumn{2}{|c|}{~~~$\displaystyle{3+ \sqrt{4g(g+\lambda)[3(1 - \epsilon f_0V_0) + g(g+\lambda)] + 9} \over 2[3(1-\epsilon f_0V_0)
+g(g+\lambda)]}$~~~}
&\multicolumn{2}{|c|}{~~~$\displaystyle{{\lambda^2\over 6}+\sqrt{\left({\lambda^2\over 6}\right)^2-2(1-\epsilon f_0V_0) \left({\lambda^2\over 6}\right)+1}}$~~~}
\\[1em]
\hline 
&\multicolumn{2}{|c|}{} &  \multicolumn{2}{|c|}{}\\[-.5em]
$p$
&\multicolumn{2}{|c|}{$\displaystyle{2\over 3}\left(1+{g\over \lambda}\right)$}
&\multicolumn{2}{|c|}{$\displaystyle{2\gamma_0\over \lambda^2}$}
\\[1em]
\hline 
\cline{4-5}
&\multicolumn{2}{|c|}{} &  \multicolumn{2}{|c|}{}\\[-.6em]
&  \multicolumn{2}{|c|}{$g>g_{\rm cr}$} & \multicolumn{2}{|c|}{$g<g_{\rm cr}$} 
    \\[.2em]
 \cline{2-5}
 existence && &&\\[-.6em]
&~~~$f_0V_0<1+{1\over 3}g(g+\lambda)$~~~&   ---  &~~~~~~~~~~~~~---~~~~~~~~~~~~~
&~$\lambda^2>\lambda_+^2$ or $\lambda^2<\lambda_-^2$~
\\[.2em] 
\hline  
&\multicolumn{2}{|c|}{} &  \multicolumn{2}{|c|}{}\\[-.6em]
acceleration&\multicolumn{2}{|c|}{$g>{\lambda\over 2}$} & \multicolumn{2}{|c|}{$\lambda^2<\lambda_{\rm cr}^2$}
\\[.4em]
\cline{4-5}
\hline 
&\multicolumn{2}{|c|}{} &  \multicolumn{2}{|c|}{}\\[-.6em]
stability&\multicolumn{2}{|c|}{$g>g_{\rm cr}$} &\multicolumn{2}{|c|}{$g<g_{\rm cr}$} 
\\[.4em]
\hline 
\hline
&\multicolumn{2}{|c|}{} &  \multicolumn{2}{|c|}{}\\[-.6em]
$w_\phi$&\multicolumn{2}{|c|}{$\displaystyle{-1+{3\gamma_0\over 3\gamma_0+g(g+\lambda)}}$}&\multicolumn{2}{|c|}{$
\displaystyle{-1+{\lambda^2\over 3\gamma_0}}$}
\\[1em]
\hline
&\multicolumn{2}{|c|}{} &  \multicolumn{2}{|c|}{}\\[-.6em]
$w_{\rm eff}$&\multicolumn{2}{|c|}{$\displaystyle{-{g\over g+\lambda}}$}&\multicolumn{2}{|c|}{$\displaystyle{-1+{\lambda^2\over 3\gamma_0}}$}
\\[1em]
\hline
&\multicolumn{2}{|c|}{} &  \multicolumn{2}{|c|}{}\\[-.6em]
$\Omega_{\rm m}$&\multicolumn{2}{|c|}{$\displaystyle{\lambda(g+\lambda)-3\gamma_0\over (g+\lambda)^2} $}&\multicolumn{2}{|c|}{$0$}
\\[1em]
\hline
&\multicolumn{2}{|c|}{} &  \multicolumn{2}{|c|}{}\\[-.6em]
$\Omega_{\phi}$&\multicolumn{2}{|c|}{$\displaystyle{g(g+\lambda)+3\gamma_0\over (g+\lambda)^2}$}&\multicolumn{2}{|c|}{$1$}
\\[1em]
\hline
 \end{tabular}
    \caption{Two analytic solutions with $\mu=-\lambda$ in the D-BIonic and DBI theories with $f=\epsilon f_0 e^{-\mu\phi}$,
    $V=V_0 e^{-\lambda\phi}$, and $A=e^{g\phi}$.
    $g_{\rm cr}$, $\lambda_{\rm cr}$ and $\lambda_\pm$  are defined in the text.
    The case I gives a scaling solution, in which the ratio of matter 
    energy density to the scalar field energy density is constant, while the scalar field energy becomes dominant in the case II solution.
     For the accelerating universe, the existence condition coincides with the stability condition, which will be analysed in Sec \ref{stability}.}
\label{Table I}
\end{center}
\end{table}
\begin{figure}[h]
	\includegraphics[width=5cm]{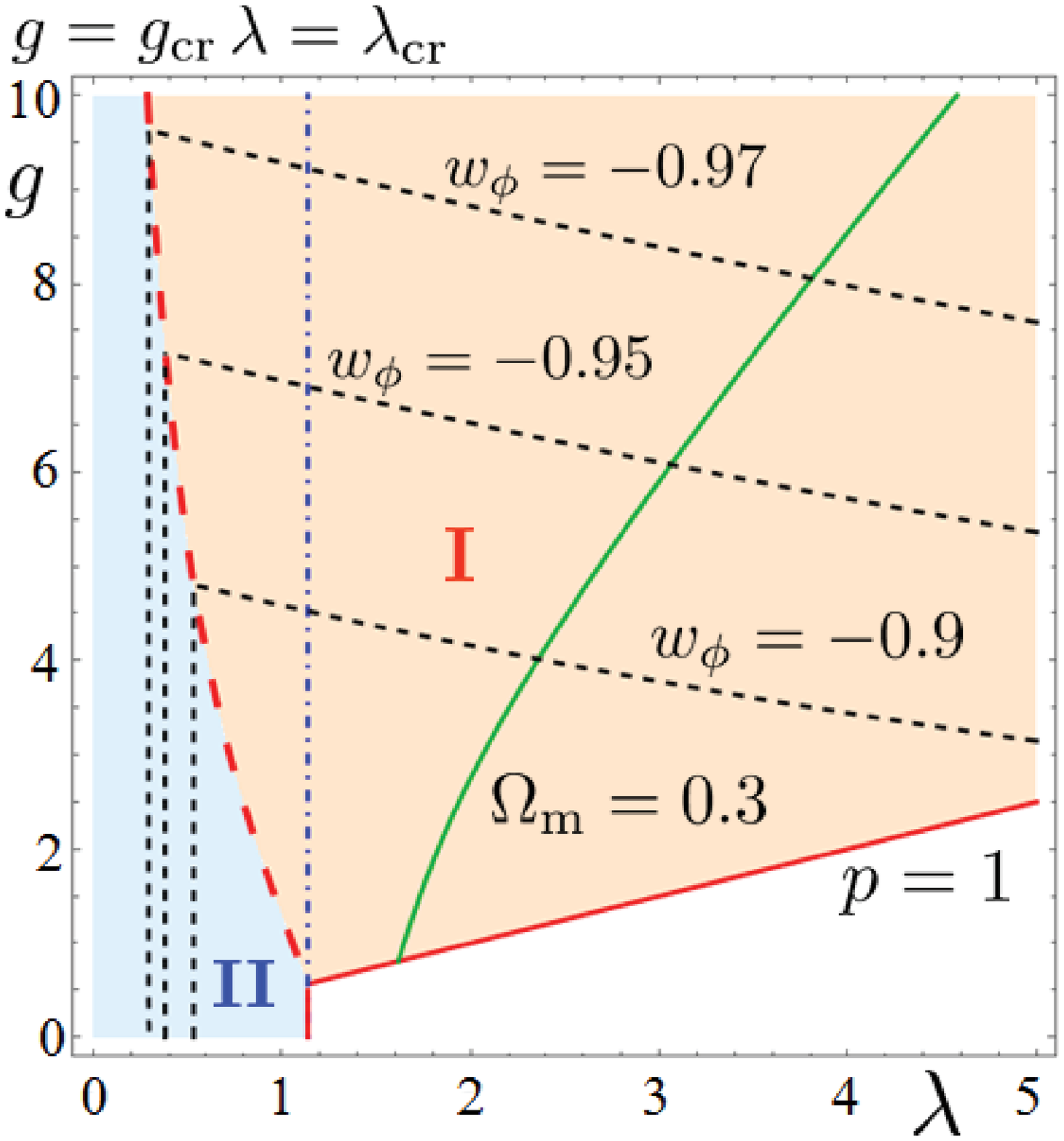} ~~
	\includegraphics[width=5cm]{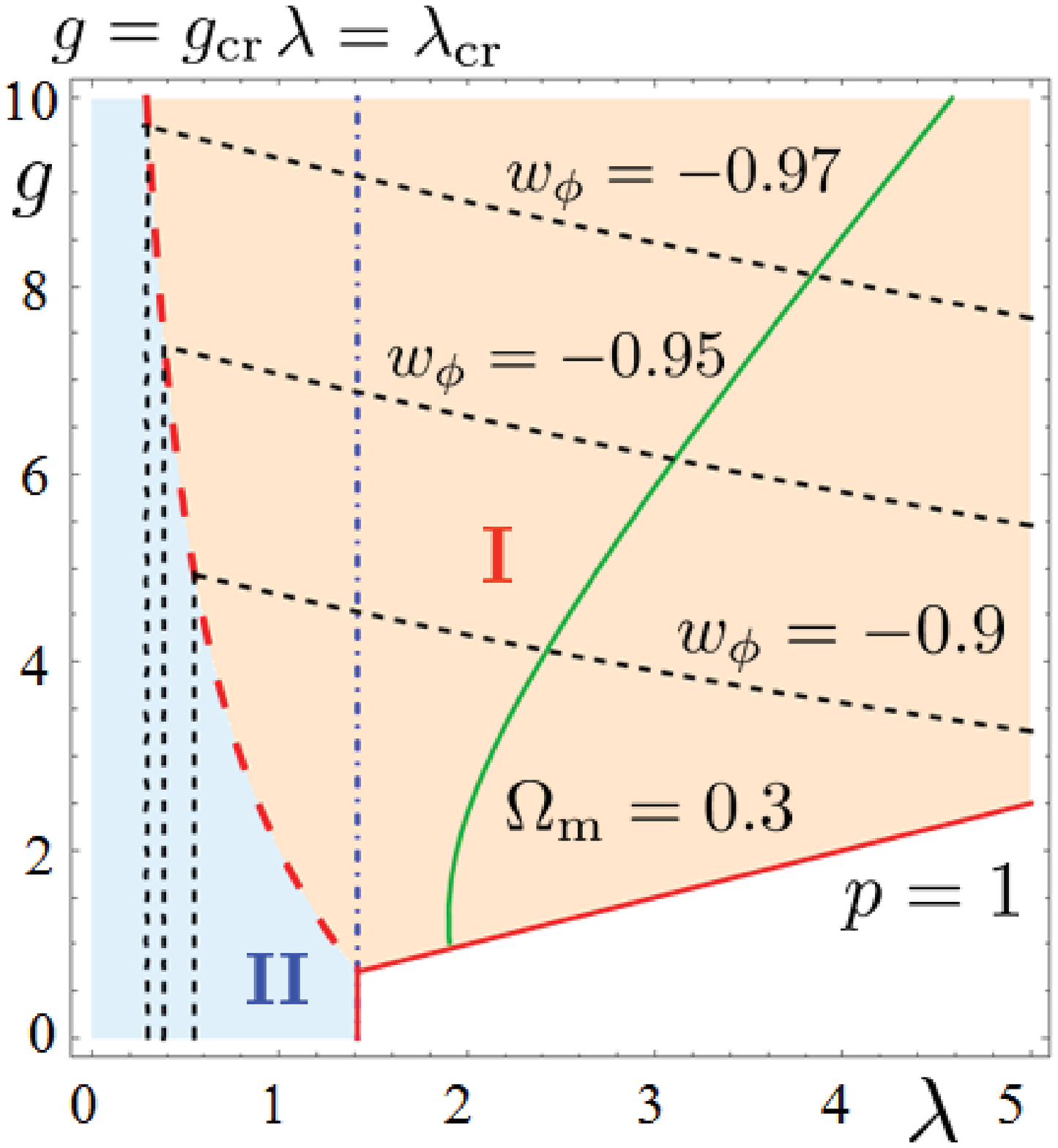} ~~
	\includegraphics[width=5cm]{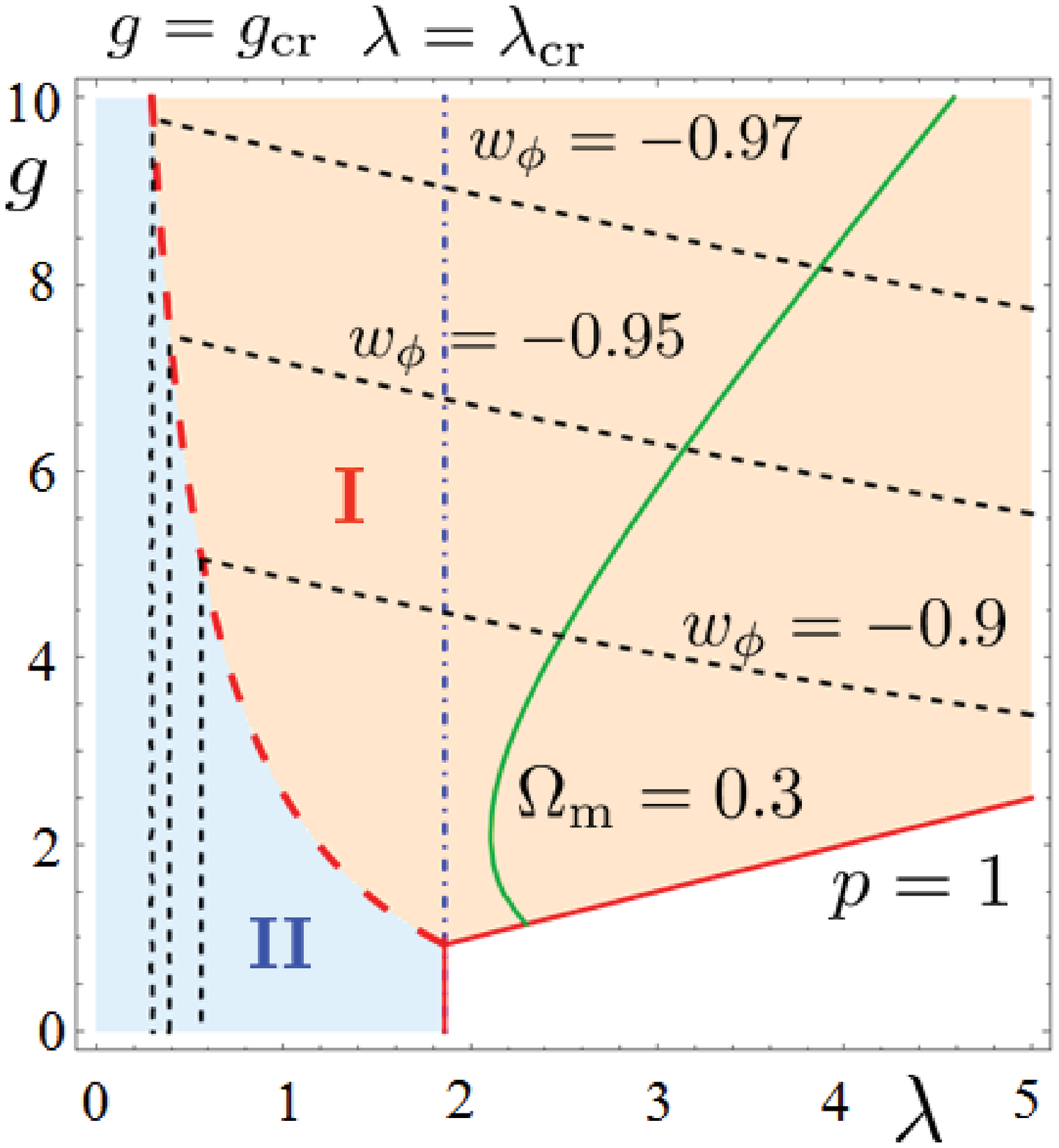}
	\\
	(a)\hskip 5cm  (b) \hskip 5cm  (c) \\
\caption{The existence regions of two accelerating 
solutions in the parameter space ($\lambda, g$) for (a) the D-BIonic ($\epsilon=-1$) , (b) the canonical kinetic term ($\epsilon=0$) and (c) the DBI  ($\epsilon=1$) .
We set $f_0V_0=1$.
The light orange and light blue  regions correspond to the solutions of the case I and II, respectively. The red dashed curve denotes $g=g_{\rm cr}$, 
while the blue dot-dashed line shows $\lambda=\lambda_{\rm cr}$. 
The green curve gives $\Omega_{\rm m}=0.3$, while the black dashed lines denote $w_{\phi}=-0.97, -0.95$ and $-0.9$, respectively  from the above. The red solid lines denote $p=1$.} 
\label{F1}
\end{figure}
\end{widetext}

If there is no matter coupling with the scalar field ($g=0$), the solution {\bf II} will 
be realised. When there exists the coupling ($g\neq 0$), there are two solutions in the range of $\lambda < \lambda_{\rm cr}$. 
The question is which asymptotic solution 
is found,  {\bf I} or  {\bf II} ?
We expect the case with the larger power exponent of the cosmic expansion $p$
will be realised \cite{FujiiMaeda}. 
For the solution  {\bf II}, the power exponent $p_{\rm II}$ is given by
\begin{equation*}
p_{\rm II}={1\over 3}\left[
1+\sqrt{1-2(1-\epsilon f_0V_0)\left({6\over \lambda^2}\right)+\left({6\over \lambda^2}\right)^2}\right]
\,,
\end{equation*}
which depends on $\epsilon f_0V_0$ and $\lambda$,
whereas for the solution {\bf I} is
\begin{equation*}
p_{\rm I}= \frac{2}{3} \left( 1 + \frac{g}{\lambda}\right)
\,,
\end{equation*}
which is fixed by $\lambda$ and $g$.
So our conjecture is that 
if $p_{\rm II} > p_{\rm I}$, then matter contribution is ignored, 
which is a usual quintessence model with the DBI or D-BIonic kinetic term,
while when $p_{\rm II} < p_{\rm I}$, the existence of matter assists the 
acceleration of the cosmic expansion. Even if $\lambda$ is too large to obtain a usual 
quintessence scenario, we find the acceleration for 
\begin{equation*}
g>{\lambda\over 2}
\,.
\end{equation*}

The critical value of the coupling constant $g$ is obtained by setting 
$p_{\rm I}=p_{\rm II}$, giving $g=g_{\rm cr}$ 
with 
\begin{equation*}
g_{\rm cr}\equiv -{\lambda\over 2}\left[1-
\sqrt{1-2(1-\epsilon f_0V_0)\left({6\over \lambda^2}\right)+\left({6\over \lambda^2}\right)^2}\right]
\,.
\end{equation*}
The critical value $g_{\rm cr}$ is the same as that for the existence obtained in the previous subsection. 
When  $g>g_{\rm cr}$, the power exponent of the solution {\bf I} 
is larger than that of the solution {\bf II}.
As we will see in the next section, the stability condition is also the same.
As a result, 
when $g>{\lambda\over 2}$ 
and $g>g_{\rm cr}$, we find the accelerated expansion of the Universe
assisted by matter fluid.

%~~\\
%\vskip 5cm

%======================================%
%<<<<<<<<<<<< SECTION IV  >>>>>>>>>>>>>>%
%======================================%
%%%%%%%%%%%%%%%%%%%%%%%%%%%%%%%%%%%%%%%%%%%%%%%%%%%%%%%%%%%%%%%%%%
%%%%%%%%%%%%%%%%%%%%%%%%%%%%%%%%%%%%%%%%%%%%%%%%%%%%%%%%%%%%%%%%%%
%%%%%%%%%%%%%%%%%%%%%%%%%%%%%%%%%%%%%%%%%%%%%%%%%%%%%%%%%%%%%%%%%%
\section{STABILITY ANALYSIS}
\label{stability}
%%%%%%%%%%%%%%%%%%%%%%%%%%%%%%%%%%%%%%%%%%%%%%%%%%%%%%%%%%%%%%%%%%
%%%%%%%%%%%%%%%%%%%%%%%%%%%%%%%%%%%%%%%%%%%%%%%%%%%%%%%%%%%%%%%%%%

In order to confirm the expectations in Sec.~\ref{expectation}, 
we need to analyse the stability of those solutions {\bf I} and {\bf II}. 
In this section we will use the dynamical system approach. 
%%%%%%%%%%%%%%%%%%%%%%%%%%%%%%%%%%%%%%%%%%%%%%%%%%%%%%%%%%%%%%%%%%
%%%%%%%%%%%%%%%%%%%%%%%%%%%%%%%%%%%%%%%%%%%%%%%%%%%%%%%%%%%%%%%%%%
\subsection{Dynamical System and Fixed points}
%%%%%%%%%%%%%%%%%%%%%%%%%%%%%%%%%%%%%%%%%%%%%%%%%%%%%%%%%%%%%%%%%%
%%%%%%%%%%%%%%%%%%%%%%%%%%%%%%%%%%%%%%%%%%%%%%%%%%%%%%%%%%%%%%%%%%
Starting from the Friedmann equation (\ref{Friedmann}), we obtain the first constraint equation on this system:
\begin{equation}
\Omega_{\rm m} = 1 - x^2 - y^2 \,, 
\label{cons1}
\end{equation}
where we introduce the following dimensionless variables;
\begin{equation*}
x \equiv \frac{\gamma}{\sqrt{3(\gamma + 1)}} \frac{\dot \phi}{H}, \quad y \equiv \frac{\sqrt{V}}{\sqrt{3} H}.
\end{equation*}
Instead of time derivatives, we use the derivatives with respect to the e-folding number, $N = \ln a$. 
We then obtain the following autonomous equations:
\begin{align}
\frac{dx}{dN} =& -\frac{3}{2}x \left[{1\over \gamma}(1-x^2)+y^2\right]
\nn
&~~~+  \frac{\sqrt{3(\gamma+1)}}{2\gamma} 
\left[\lambda y^2 - 
g  (1 - x^2 - y^2) \right]\,, 
\label{eq_x}
\\[1em]
\frac{dy}{dN} =&   \frac{3}{2}y\left[{1\over \gamma} x^2 + (1  - y^2)\right]
- \lambda \frac{\sqrt{3(\gamma+1)}}{2\gamma} x y
\,.
\label{eq_y}
\end{align}

Since the variable $\gamma$ is included in the above equations,  in order to close 
the system, we need the equation for  $\gamma$, which is given by
\begin{align}
\frac{d \gamma}{d N} &=  \frac{(\gamma -1)\sqrt{3 (\gamma + 1)} }{\gamma x} \times \nn
&\left[ - \sqrt{3(\gamma + 1)}x  - \mu x^2 +   \lambda y^2 -g (1 - x^2 -y^2)  \right] 
\,.
\label{gammaN}
\end{align}
However, note that  $\gamma$ is described as
\begin{eqnarray*}
\gamma=1+\epsilon f_0 V_0 e^{-(\lambda+\mu)\phi}{x^2\over y^2}
\,.
\end{eqnarray*}
Hence when $\mu=-\lambda$, $\gamma$ is not the independent variable.
Eqs. (\ref{eq_x}) and (\ref{eq_y}) give a closed set of the dynamical system.

By virtue of these dynamical variables, the cosmological parameters are given by
\begin{align*}
\Omega_{\phi} &= x^2 + y^2\,, \\
w_{\phi} &= \frac{x^2 - \gamma y^2}{\gamma(x^2 + y^2)}\,, \\
w_{\rm eff} &= \frac{1}{\gamma} \left( x^2 -\gamma  y^2\right)\,.
\end{align*}

We are interested in the fixed points $(x,y)=(x_0,y_0)$ with $\gamma = \gamma_0 = $ constant, which yields
 $d \gamma / d N = 0$. Since $\gamma>0$, we find the following two 
possibilities: \\
{\bf (i)} $\gamma_0 =1$, this is the same as the coupled quintessence with the conventional 
canonical kinetic term, which is not our interest. \\
{\bf (ii)} The intermediate value of $\gamma_0$, i.e.  $0<\gamma_0<1$ for the D-BIonic theory, while $1<\gamma_0<\infty$ for 
the DBI theory, 
which is obtained  from the condition such that  the square bracket in Eq.~(\ref{gammaN}) is equal to zero.
We find 
\begin{equation}
\gamma_0=-1+{1\over 3 x_0^2} \left[ \lambda y_0^2 - \mu x_0^2 -g (1 - x_0^2 -y_0^2)\right]^2 \,. 
\label{gammaiii}
\end{equation}

Since $\gamma_0=1$ does not give new solution, we will discuss only the case {\bf (ii)}.
By setting $dx/dN = 0$ and $dy/dN = 0$ with $\gamma=\gamma_0$, 
we find   fixed points ($x_0,y_0$) as shown in Table \ref{tab2}.
\begin{table}[h]
\centering
%\large
\begin{tabular}{ c | c  c | c}
\hline
\hline
 & $x_0$ & $y_0$ & solution\\[.5em] 
\hline
  &   &  & \\[-.5em]
(1) & $-1$ & 0 &{\small\bf IV}$_-$\\[1em] 
(2) & $ 1$ & 0 &{\small\bf IV}$_+$\\[.5em] 
 (3) & $-\frac{g  \sqrt{3(1+\gamma_0)}}{3  }$ & 0 &{\small\bf III}\\[1em] 
 (4) & $\frac{\lambda}{\sqrt{3(1+\gamma_0)}}$ & $\sqrt{1 - \frac{\lambda^2}{3(1+\gamma_0)}}$ &{\small\bf II} \\[1em] 
 (5) & $\frac{\sqrt{3} \gamma_0}{\sqrt{1+\gamma_0} (g   + \lambda)}$ & 
$\sqrt{\frac{3 \gamma_0  + (1+\gamma_0)g (g+\lambda)}{(1+\gamma_0)(g   + \lambda)^2}}$&{\small\bf I} \\[.5em] 
\hline
\hline
\end{tabular} 
\caption{Five fixed points. }
\label{tab2}
\end{table}

There are five fixed points, which satisfy the necessary condition of $y_0 \geq 0$, whereas $x$ can be positive or negative depending 
on the sign of $\dot \phi$.

We expect that these fixed points correspond to the (asymptotic) analytic solutions given in the previous section
 and Appendix \ref{other_solutions}. 
We shall describe each point in the following:

%%%%%%%%%%%%%%%%%%%%%%%%%%%%%%%%%%%%%%%%%%%%%%%%%%%%%%%%%%%%%%%%%%%%%%%%%%%%%%%
%%%%%%%%%%%%%%%%%%%%%%%%%%%%%%%%%%%%%%%%%%%%%%%%%%%%%%%%%%%%%%%%%%%%%%%%%%%%%%%
\subsubsection{\rm Fixed points (1) and (2)}
%%%%%%%%%%%%%%%%%%%%%%%%%%%%%%%%%%%%%%%%%%%%%%%%%%%%%%%%%%%%%%%%%%%%%%%%%%%%%%%
%%%%%%%%%%%%%%%%%%%%%%%%%%%%%%%%%%%%%%%%%%%%%%%%%%%%%%%%%%%%%%%%%%%%%%%%%%%%%%%
The 
%most 
simplest fixed points 
%is 
are given by 
\begin{align*}
(1) \qquad (x_0,y_0) &= (-1,0) \,, \nn
(2) \qquad (x_0,y_0) &= (1,0) \,. \nn
\end{align*}

From Eq.~(\ref{gammaiii}), we obtain
\begin{equation*}
\gamma_0 = \frac{\mu^2 -3}{3} \,,
\end{equation*}
and the cosmological parameters as
\begin{align*}
\Omega_{\phi} &= 1 \,, \\
\Omega_{\rm m} &= 0 \,, \\
w_{\phi} &= \frac{1}{\gamma_0} \,, \\
w_{\rm eff} &= \frac{1}{\gamma_0} \,.
\end{align*}
Thus, the fixed points (1) and (2) correspond to the asymptotic solution {\bf IV}$_\pm$ given in Appendix \ref{other_solutions}. 
Since this solution does not give an accelerating Universe, we will not analyse the stability, 
although we expect it is unstable unless $V_0=0$ \cite{Copeland}.

%%%%%%%%%%%%%%%%%%%%%%%%%%%%%%%%%%%%%%%%%%%%%%%%%%%%%%%%%%%%%%%%%%%%%%%%%%%%%%%
%%%%%%%%%%%%%%%%%%%%%%%%%%%%%%%%%%%%%%%%%%%%%%%%%%%%%%%%%%%%%%%%%%%%%%%%%%%%%%%
\subsubsection{\rm Fixed point (3)}
%%%%%%%%%%%%%%%%%%%%%%%%%%%%%%%%%%%%%%%%%%%%%%%%%%%%%%%%%%%%%%%%%%%%%%%%%%%%%%%
%%%%%%%%%%%%%%%%%%%%%%%%%%%%%%%%%%%%%%%%%%%%%%%%%%%%%%%%%%%%%%%%%%%%%%%%%%%%%%%
Next simple fixed point is found as
\begin{align*}
(3) \qquad (x_0,y_0) &= \left(-\frac{ g  \sqrt{3(1+\gamma_0)}}{3},0\right) \,.\nn
\end{align*}

From Eq.~(\ref{gammaiii}), we obtain
\begin{equation*}
\gamma_0 = \frac{g(\mu - g)}{3 - g(\mu - g)} \,,
\end{equation*}
and the cosmological parameters as
\begin{align*}
\Omega_{\phi} &= \frac{g^2 (1 + \gamma_0)}{3} \,, \\
\Omega_{\rm m} &=  \frac{3 - g^2 (1 + \gamma_0)}{3} \,, \\
w_{\phi} &= \frac{1}{\gamma_0} \,, \\
w_{\rm eff} &= \frac{g^2 (1 + \gamma_0)}{3 \gamma_0} \,.
\end{align*}
Then, the fixed points (3) corresponds to the asymptotic solution {\bf III} discussed in Appendix \ref{other_solutions}.
Since this solution does not give an accelerating Universe either, we will not analyse the stability.
(In this case, we also expect the fixed point is unstable \cite{Copeland}).

%%%%%%%%%%%%%%%%%%%%%%%%%%%%%%%%%%%%%%%%%%%%%%%%%%%%%%%%%%%%%%%%%%%%%%%%%%%%%%%%%%%%%%%%%%%%%%%%%%%%%%%%
\subsubsection{\rm Fixed point (4)}
%%%%%%%%%%%%%%%%%%%%%%%%%%%%%%%%%%%%%%%%%%%%%%%%%%%%%%%%%%%%%%%%%%%%%%%%%%%%%%%%%%%%%%%%%%%%%%%%%%%%%%%%
One interesting fixed point is given by 
\begin{align*}
(4) \qquad (x_0,y_0) = \left(\frac{\lambda}{\sqrt{3(1+\gamma_0)}},\sqrt{1 - \frac{\lambda^2}{3(1+\gamma_0)}}\right) \,.
\end{align*}

From  Eq. (\ref{gammaiii}),  we find  
\begin{equation*}
\mu = - \lambda \,.
\end{equation*}
By using the definition of $\gamma$ and the above relation, we have
\begin{equation}
\gamma = \frac{\epsilon f_0 V_0 x^2 + y^2}{y^2} \,. 
\label{cons2}
\end{equation}
Substituting $x_0$ and $y_0$ of the fixed point (4) in Eq. (\ref{cons2}), we obtain the equation for $\gamma_0$ as
\begin{equation*}
3\gamma_0^2-\lambda^2\gamma_0+(1-\epsilon f_0V_0)\lambda^2-3=0 
\,.
\end{equation*}
This is Eq.~(\ref{gammaC2}) of the solution {\bf II} as we expect. $\gamma_0$  is given by
\begin{equation*}
\gamma_0 =
{\lambda^2\over 6}\left[1 + \sqrt{1-2(1-\epsilon f_0V_0)\left({ 6\over \lambda^2} \right)
+\left({ 6\over \lambda^2}\right)^2}\right]
\,.
\end{equation*}
Only the larger root of the solutions is chosen because it 
%satisfies 
gives the accelerated expansion ($p > 1$).

The cosmological parameters  also confirm that there is no contribution from matter density at the fixed point (4).
\begin{align*}
\Omega_{\phi} &= 1 \,, \\
\Omega_{\rm m} &=  0 \,, \\
w_{\phi} &= -1 + \frac{\lambda^2}{3\gamma_0} \,, \\
w_{\rm eff} &= -1 + \frac{\lambda^2}{3\gamma_0} \,.
\end{align*}

The fixed point (4) is the same as the point (C4) in Ref. \cite{Copeland}.  
%%%%%%%%%%%%%%%%%%%%%%%%%%%%%%%%%%%%%%%%%%%%%%%%%%%%%%%%%%%%%%%%%%%%%%%%%%%%%%%
%%%%%%%%%%%%%%%%%%%%%%%%%%%%%%%%%%%%%%%%%%%%%%%%%%%%%%%%%%%%%%%%%%%%%%%%%%%%%%%
\subsubsection{\rm Fixed point (5)}
%%%%%%%%%%%%%%%%%%%%%%%%%%%%%%%%%%%%%%%%%%%%%%%%%%%%%%%%%%%%%%%%%%%%%%%%%%%%%%%
%%%%%%%%%%%%%%%%%%%%%%%%%%%%%%%%%%%%%%%%%%%%%%%%%%%%%%%%%%%%%%%%%%%%%%%%%%%%%%%
The last fixed point gives another interesting solution:
%\begin{widetext}
\begin{align*}
&(5) \qquad \nn
&(x_0,y_0) = 
\left(\frac{\sqrt{3} \gamma_0}{\sqrt{1+\gamma_0} (g  + \lambda)},
 \sqrt{\frac{3\gamma_0 +  (1+\gamma_0)g(g+\lambda)}{(1+\gamma_0)(g   + \lambda)^2}}\right) \,.
\end{align*}
%\end{widetext}

In the similar way as the fixed point (4), 
we also have $\mu = -\lambda$ in this solution from Eq.~(\ref{gammaiii}),
and then
we obtain from Eq.~(\ref{cons2}) the quadratic equation for $\gamma_0$ as 
\begin{equation*}
[3(1-\epsilon f_0V_0)
+g(g+\lambda)]\gamma_0^2
-3\gamma_0
-g(g+\lambda)=0
\,.
\end{equation*}
This is the sams as Eq.~(\ref{gamma_I}) which is found  in the solution {\bf I}. The solution is. 
\begin{equation*}
\gamma_0 = \frac{3 + \sqrt{4g(g+\lambda)[3(1 - \epsilon  f_0 V_0 ) + g(g +\lambda)]+ 9}}{2 (3(1 - \epsilon  f_0 V_0 ) 
+ g(g +\lambda))} \,.
\end{equation*}
We choose only larger root because of the same reason as the previous subsection.

The cosmological parameters are
\begin{align}
\Omega_{\phi} &=  \frac{g (g +\lambda) + 3\gamma_0}{(g+\lambda)^2} \,, \\
\Omega_{\rm m} &= \frac{\lambda (g+\lambda) - 3\gamma_0}{(g+\lambda)^2} \,, \\
w_{\phi} &= - \frac{g (g+\lambda)}{3\gamma_0 + g(g+\lambda)} \,, 
\label{wphiD} \\
w_{\rm eff} &= - \frac{g}{(g+\lambda)} \,. 
\label{weffD}
\end{align}
This is the scaling solution as we have seen in the solution {\bf I}. 

The fixed point (5) is the extension of the fixed point (C5) in Ref. \cite{Copeland}.
In fact, it is the same as 
the fixed point (C5) when there is no coupling ($g = 0$).

Next, we will analyse the stability of the fixed points (4) and (5) (the solutions {\bf I} and {\bf II}).

%%%%%%%%%%%%%%%%%%%%%%%%%%%%%%%%%%%%%%%%%%%%%%%%%%%%%%%%%%%%%%%%%%%%%%%%%%%%%%%%%%%%%%%%%%%%%
%%%%%%%%%%%%%%%%%%%%%%%%%%%%%%%%%%%%%%%%%%%%%%%%%%%%%%%%%%%%%%%%%%%%%%%%%%%%%%%%%%%%%%%%%%%%%
\subsection{Linear stability}
%%%%%%%%%%%%%%%%%%%%%%%%%%%%%%%%%%%%%%%%%%%%%%%%%%%%%%%%%%%%%%%%%%%%%%%%%%%%%%%%%%%%%%%%%%%%%
%%%%%%%%%%%%%%%%%%%%%%%%%%%%%%%%%%%%%%%%%%%%%%%%%%%%%%%%%%%%%%%%%%%%%%%%%%%%%%%%%%%%%%%%%%%%%

Substituting Eq.~(\ref{cons2}) into Eqs.~(\ref{eq_x}) and  (\ref{eq_y}) , we obtain the autonomous system only for $x$ and $y$:
\begin{align*}
\frac{dx}{dN} = F(x,y) \,, \quad \frac{dy}{dN} = G(x,y) \,.
\end{align*}
Considering linear perturbations 
\begin{equation*}
\begin{pmatrix} x \\  y \end{pmatrix} =
\begin{pmatrix} x_0+\delta x \\ y_0+\delta y \end{pmatrix} \,,
\end{equation*}
we find
\begin{equation*}
\frac{d}{dN} \begin{pmatrix} \delta x \\ \delta y \end{pmatrix} =
 \mathcal{M}_0 \begin{pmatrix} \delta x \\ \delta y \end{pmatrix} \,,
\end{equation*}
where 
\begin{equation*}
\mathcal{M}_0 = \begin{pmatrix} \frac{\partial F}{\partial x}\Big{|}_0 && \frac{\partial F}{\partial y}\Big{|}_0 \\[.5em]
\frac{\partial G}{\partial x}\Big{|}_0 && \frac{\partial G}{\partial y}\Big{|}_0 \end{pmatrix} \,.
\end{equation*}
%The 
Each component of the matrix $\mathcal{M}$ is given by 
\begin{widetext}
\begin{align*}
\frac{\partial F}{\partial x}\Big{|}_0=& \frac{1}{2x_0 \gamma_0^2 \sqrt{1+ \gamma_0}} 
\Big{\{}3x_0 \sqrt{1+\gamma_0} \left[\gamma_0-2+( \gamma_0+2)x_0^2 - \gamma_0^2 y_0^2 \right]   \nn 
&  +\sqrt{3}\left[(\gamma_0-1)(\gamma_0+2)(g-(g+\lambda) y_0^2)+(\gamma_0^2+\gamma_0+2)x_0^2\right]
\Big{\}} \,, \\
\frac{\partial F}{\partial y}\Big{|}_0 =& \frac{1}{2y_0 \gamma_0^2 \sqrt{1+ \gamma_0}} 
\Big{\{}6x_0 \sqrt{1+\gamma_0} \left[(\gamma_0-1)(x_0^2-1)-\gamma_0^2 y_0^2\right]  \nn
&   + \sqrt{3}\left[g(\gamma_0-1)(\gamma_0+2)(x_0^2-1)+(g+\lambda)(3\gamma_0^2+3\gamma_0-2)y_0^2\right]
\Big{\}} \,, \\
\frac{\partial G}{\partial x}\Big{|}_0 =& \frac{y_0}{\gamma_0^2\sqrt{1+\gamma_0}}
\left(3x_0\sqrt{1+\gamma_0}  - \sqrt{3}\lambda\right) \,, \\
\frac{\partial G}{\partial y}\Big{|}_0 =& - \frac{3}{2\gamma_0^2 } \left[(2-3\gamma_0)x_0^2+3\gamma_0^2 y_0^2-\gamma_0^2\right]
- {\sqrt{3}\lambda(\gamma_0^2+\gamma_0-1)\over \gamma_0^2\sqrt{1+\gamma_0}} x_0
 \,.
\end{align*}
\end{widetext}

Setting 
\begin{align*}
\delta x, \delta y \propto e^{\omega N}\,,
\end{align*}
we find the quadratic equation for the eigenvalues $\omega$ of the matrix $\mathcal{M}_0$.
If both eigenvalues are negative (or real parts are negative for complex eigenvalues), 
 the fixed point is stable against linear perturbations.

%%%%%%%%%%%%%%%%%%%%%%%%%%%%%%%%%%%%%%%%%%%%%%%%%%%%%%%%%%%%%%%%%%%%%%%%%%%%%%%%%%%%%%%%%%%
%%%%%%%%%%%%%%%%%%%%%%%%%%%%%%%%%%%%%%%%%%%%%%%%%%%%%%%%%%%%%%%%%%%%%%%%%%%%%%%%%%%%%%%%%%%
\subsubsection{\rm Fixed point (4)}
%%%%%%%%%%%%%%%%%%%%%%%%%%%%%%%%%%%%%%%%%%%%%%%%%%%%%%%%%%%%%%%%%%%%%%%%%%%%%%%%%%%%%%%%%%%
%%%%%%%%%%%%%%%%%%%%%%%%%%%%%%%%%%%%%%%%%%%%%%%%%%%%%%%%%%%%%%%%%%%%%%%%%%%%%%%%%%%%%%%%%%%

For the fixed point (4),  we find two real eigenvalues  of the matrix $\mathcal{M}_0$ as 
\begin{equation*}
\omega_1 = -3 + \frac{\lambda^2}{2\gamma_0} \,, ~~\omega_2 =-3  + \frac{\lambda (g  + \lambda)}{\gamma_0} \,.
\end{equation*}
In order for the fixed point to be stable, both eigenvalues must be negative, which condition requires
\begin{equation*}
 \lambda^2 < 6 \gamma_0  \quad {\rm and} \quad  \lambda (g + \lambda)  < 3\gamma_0  \,.
\end{equation*}
The first condition is always true when we choose $\gamma_0^{(+)}$ for the accelerating universe solution. 
While, the second condition gives
\begin{equation*}
g < g_{\rm cr} \,.
\end{equation*}
This confirms our anticipation in the previous section that we must impose the condition 
$g < g_{\rm cr}$ in order to obtain the stable solution
of  {\bf II}.
The solution {\bf II} in the light blue region in Fig. \ref{F1} is stable.

%%%%%%%%%%%%%%%%%%%%%%%%%%%%%%%%%%%%%%%%%%%%%%%%%%%%%%%%%%%%%%%%%%%%%%%%%%%%%%%%%%%%%%%%%%%
%%%%%%%%%%%%%%%%%%%%%%%%%%%%%%%%%%%%%%%%%%%%%%%%%%%%%%%%%%%%%%%%%%%%%%%%%%%%%%%%%%%%%%%%%%%
\subsubsection{\rm Fixed point (5)}
%%%%%%%%%%%%%%%%%%%%%%%%%%%%%%%%%%%%%%%%%%%%%%%%%%%%%%%%%%%%%%%%%%%%%%%%%%%%%%%%%%%%%%%%%%%
%%%%%%%%%%%%%%%%%%%%%%%%%%%%%%%%%%%%%%%%%%%%%%%%%%%%%%%%%%%%%%%%%%%%%%%%%%%%%%%%%%%%%%%%%%%

In the similar way as the fixed point (4), the eigenvalues of the fixed point (5) are obtained as
%\begin{widetext}
\begin{equation*}
\omega_\pm = \frac{-3 \gamma_0^{3/2} (2g+ \lambda) \pm \sqrt{\mathcal{D}}}{4\gamma_0^{3/2}(g+\lambda)} \,.
\end{equation*}
with
\begin{eqnarray*}
&&
\mathcal{D}\equiv 9\gamma_0^3 (2g+\lambda)^2 
\nn
&&~~~
+24[3\gamma_0+2g(g+\lambda)][3\gamma_0-\lambda(g+\lambda)]]
~~~~~
\end{eqnarray*}
%\end{widetext}
From the condition (\ref{con_density})  of the solution {\bf I} , 
\begin{eqnarray*}
\mathcal{D}\leq 9\gamma_0^3 (2g+\lambda)^2 \,.
\end{eqnarray*}
Hence   if $\mathcal{D}\geq 0$,  $\sqrt{\mathcal{D}}$ is always   smaller than 
$3 \gamma_0^{3/2} (2g+ \lambda)$. Thus, we find $\omega_\pm \leq 0$. Consequently the solution is stable.
When $\mathcal{D}<0$, the square root term $\sqrt{\mathcal{D}}$ is pure imaginary, then 
${\rm Re}(\omega_\pm)<0$. It guarantees that the solution is again stable (with spiral trajectories). 
We also find the marginal stable condition $\omega_+=0$ when $3\gamma_0-\lambda(g+\lambda)=0$, which corresponds to 
$\rho_0=0$, i.e., $g=g_{\rm cr}$. This gives the boundary of the stable region in the parameter space.
As a result, the solution {\bf I} in the light orange region in Fig. \ref{F1} is always stable.

%%%%%%%%%%%%%%%%%%%%%%%%%%%%%%%%%%%%%%%%%%%%%%%%%%%%%%%%%%%%%%%%%%%%%%%%%%%%%%%%%%%%%%%%%%%%%
%%%%%%%%%%%%%%%%%%%%%%%%%%%%%%%%%%%%%%%%%%%%%%%%%%%%%%%%%%%%%%%%%%%%%%%%%%%%%%%%%%%%%%%%%%%%%
\subsection{Non-linear stability}
%%%%%%%%%%%%%%%%%%%%%%%%%%%%%%%%%%%%%%%%%%%%%%%%%%%%%%%%%%%%%%%%%%%%%%%%%%%%%%%%%%%%%%%%%%%%%
%%%%%%%%%%%%%%%%%%%%%%%%%%%%%%%%%%%%%%%%%%%%%%%%%%%%%%%%%%%%%%%%%%%%%%%%%%%%%%%%%%%%%%%%%%%%%

In order to see whether the stable solutions are natural or not, we have to study not only the linear stability
but also the global (or non-linear) stability. 
Here we solve the basic equations numerically and show those fixed points 
are globally stable.

\begin{figure}[htb]
	\includegraphics[width=8.5cm]{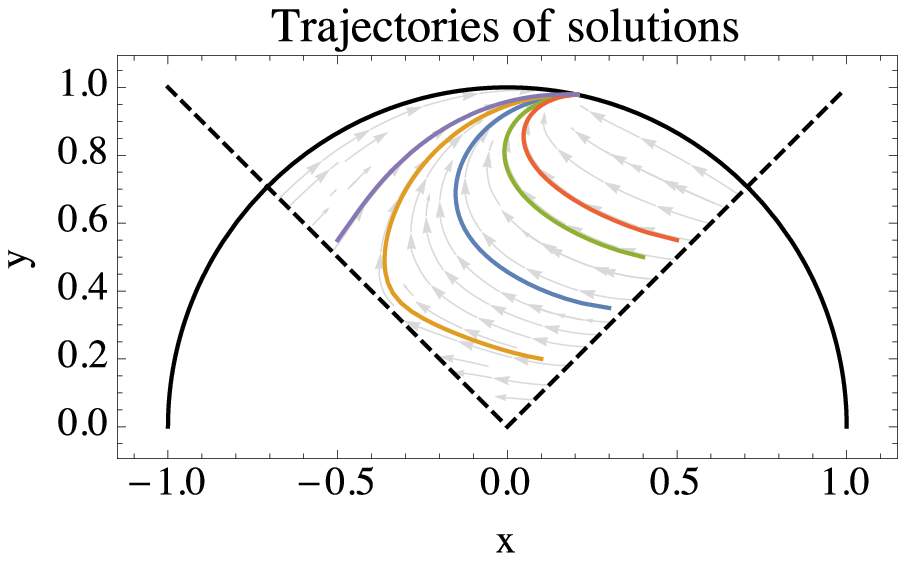} \\
	(a) $\lambda=0.5$\\[1em]
	%\begin{center} \qquad (a) \end{center}
	\includegraphics[width=8.5cm]{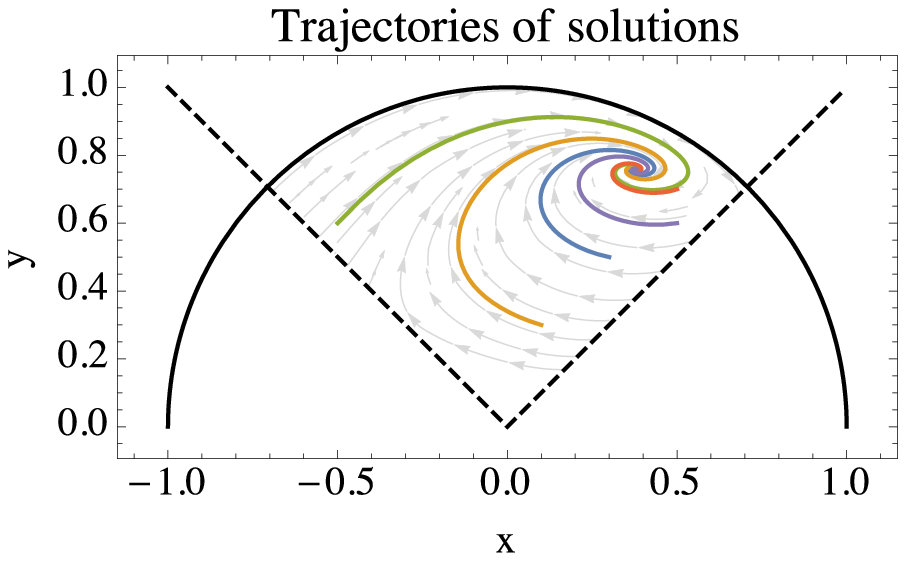}\\
	(b) $\lambda=1.6$
	\caption{The trajectries of numerical solutions of the autonomous equations and fixed points for the D-BIonic gravity theory.  
	The black thick curve denotes the limiting condition of $\Omega_{\rm m} = 0$   
and the black dashed lines correspond to the boundaries of $\gamma = 0$. 
We choose $g = 1$, and $\epsilon f_0 V_0 = -1$. 
(a) The top figure shows that the trajectories of the solutions converge to the fixed point $(x_0,y_0)=(0.206,0.978)$, which is 
the fixed point (4) with $\Omega_{\rm m}=0$ (the solution {\bf II}).
(b) For the bottom figure, the trajectories  converge to another stable fixed point  $(x_0,y_0)=(0.378,0.758)$ which is 
the fixed point (5) with $\Omega_{\rm m}=0.28 $ (the solution {\bf I}).}
\label{phasespace}
\end{figure}

In Fig {\ref{phasespace}}, we present some examples for the D-BIonic theory.
We choose $g=1$ and $\epsilon f_0 V_0 = -1$, giving 
 $\lambda_{\rm cr} = 1.14$. 
In Fig {\ref{phasespace}} (a), we show the case of  $\lambda=0.5$, which is $g < g_{\rm cr}=5.23$. 
The trajectories of the numerical solutions show that they converge to the stable fixed point
$(x_0,y_0)=(0.206,0.978)$ with $\gamma_0=0.96$, which is the fixed point (4)  (the solution {\bf II}). 
The black thick curve denotes the limiting condition of $\Omega_{\rm m} = 0$ found by (\ref{cons1})   
and the black dashed lines correspond to the boundaries of $\gamma = 0$ given by the definition (\ref{cons2}). 

In Fig {\ref{phasespace}} (b), we depict the trajectories of the solutions for the case of   $\lambda=1.6$, 
which satisfies $\lambda > \lambda_{\rm cr}$ as well as $\lambda < 2 g$.
%is $g > g_{\rm cr}=\red{??}$. 
The trajectories converge to another stable fixed point 
$(x_0,y_0)=(0.378,0.758)$ with $\gamma_0=0.75$, which is the fixed point (5)  (the solution {\bf I}). 
In this case we find the asymptotic value of $\Omega_{\rm m}=0.28 (\neq 0)$.

We also present the time evolution of the density parameters $\Omega_{\rm m}$ and $\Omega_\phi$
for the solution {\bf I}
in Fig. \ref{evo_Omega}.
For the case {\bf II}, we expect those values approach to  $0$ and $1$.
But the case {\bf I} shows those asymptotic values are some intermediate values 
between  $0$ and $1$.  Since the present observational values show the latter case, 
if we select the case {\bf II}, we may need to fine-tune the present time to explain the observed values (the coincidence problem).
On the other hand, when we adopt the case {\bf I}, we may explain the observed values just by the 
asymptotic ones.  We need not to fine-tune the present time.

\begin{figure}[htb]
%	\includegraphics[width=6cm]{omegasolutionII.eps} \\
%	(a) $g=0.1$\\[1em]
	%\begin{center} \qquad (a) \end{center}
	\includegraphics[width=7cm]{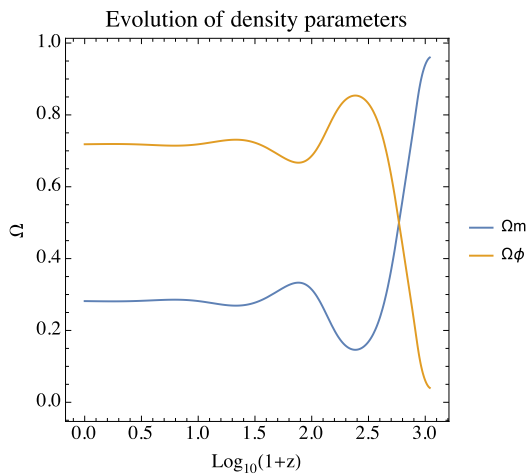}\\
%	(b) $g=3$
	\caption{The time evolution of the density parameters $\Omega_{\rm m}$ and $\Omega_\phi$ in terms of the red shift $z$. 
We choose $\lambda = - \mu = 1.6$, $\epsilon f_0 V_0 = -1$ and 
 $g=1$ just as Fig. \ref{phasespace} (b). 
 The values approach $(\Omega_{\rm m},\Omega_\phi)=(0.28, 0.72)$ at the fixed point (5) 
 (the solution {\bf I}).}
\label{evo_Omega}
\end{figure}

%======================================%
%<<<<<<<<<<<< SECTION V  >>>>>>>>>>>>>>%
%======================================%
%%%%%%%%%%%%%%%%%%%%%%%%%%%%%%%%%%%%%%%%%%%%%%%%%%%
%%%%%%%%%%%%%%%%%%%%%%%%%%%%%%%%%%%%%%%%%%%%%%%%%%%
%%%%%%%%%%%%%%%%%%%%%%%%%%%%%%%%%%%%%%%%%%%%%%%%%%%
\section{Observational constraints} 
\label{observations}
%%%%%%%%%%%%%%%%%%%%%%%%%%%%%%%%%%%%%%%%%%%%%%%%%%%
%%%%%%%%%%%%%%%%%%%%%%%%%%%%%%%%%%%%%%%%%%%%%%%%%%%

In this section, we study whether we can explain the coincidence problem  of dark energy and dark matter or not. 
We assume that the Universe at present is described by the scaling solution {\bf I}.

 From observations,  we have the constraints on the cosmological parameters as 
 $w_{\rm DE} = -0.97\pm 0.05$, $\Omega_{\rm DE} = 0.692 \pm 0.012$, and $\Omega_{\rm CDM + B} = 0.308 \pm 0.012$ \cite{Aubourg}.
 
As for the constraint on the coupling constant $g$,
it was shown  $|g| \lsim 0.13$ from the CMB observation \cite{Amendola2}. 
Although our model is different from theirs (the type {\bf II} tracking solution with the canonical kinetic term),
we expect the coupling constant is not so large.
Here we then assume the upper bound value on $g$, i.e., $g\approx 0.1$.
From the EOS parameter of dark energy,
it gives the strong constraint on $\gamma_0$ as follows:
 $\gamma_0$ of the solution {\bf I} is given by 
 \begin{eqnarray}
 \gamma_0=-{g(g+\lambda)(1+w_\phi)\over 3w_\phi}
 \,.
 \label{gamma_DE_EOS}
 \end{eqnarray}
 Since the acceleration condition is $\lambda<2g$, 
 we find 
  \begin{eqnarray}
 \gamma_0\lsim 3\times 10^{-4}\ll 1
 \label{gamma_obs_con}
 \,.
 \end{eqnarray}
Obviously, $\epsilon$ must be negative ($\epsilon = -1$), and then 
only the D-BIonic theory can provide such a solution.
The condition (\ref{gamma_obs_con}) yields very large 
value of $f_0 V_0$.
In fact, assuming $f_0 V_0\gg 1$, we find from Eqs.~(\ref{gamma_I}) and (\ref{gamma_DE_EOS})
 \begin{eqnarray}
 f_0 V_0\sim  {3w_\phi^2\over g(g+\lambda)(1+w_\phi)^2} \gsim 10^5
  \label{f0V0_obs_con}
 \,.
 \end{eqnarray}
Note that there is no upper bound on $f_0 V_0$ in the D-BIonic theory unlike Eq.~(\ref{con_DBI2}) in the DBI case.

%Then, in order to obtain the scaling solution {\bf I} by using Eq. (\ref{weffD}), we find
%\begin{align}
%w_{\rm eff} &\approx -0.67 \,, \\
%\lambda &\approx 0.049 \,.
%\end{align}
%These results lead to the fact that we require
%\begin{align}
%\gamma_0 &\approx 1.54 \times 10^{-4} \,, \\
%\epsilon f_0 V_0 &\approx - 2.17 \times 10^5 \,.
%\end{align}
%The above parameters are obtained from Eqs. (\ref{wphiD}) and (\ref{gamma_I}).
%Obviously, $\epsilon$ must be negative ($\epsilon = -1$), while the value of $f_0 V_0$ is of the order $10^5$. This means only D
%-BIonic model can satisfy because $\epsilon$ is negative and there is no upper bound on $f_0 V_0$ as Eq. (\ref{con_DBI2}) in DBI case. 
%(the upper bound forces $f_0 V_0$ of DBI theory to be $f_0 V_0 < 1.00497$).

In addition to the above cosmological constraint, we have another constraint for the screening at smaller scale.
According to \cite{Burrage}, the solar system constraints on the D-BIonic 
%gravity 
theory is 
\begin{equation*}
\sqrt{g} \Lambda \lesssim 4 \times 10^{-5} \, {\rm eV} \,,
\end{equation*}
where $\Lambda$ is a typical mass scale of the screening and defined by the action (\ref{action0}) in Appendix A.
Comparing our exponential form of $f(\phi)$ to the original paper of the D-BIonic theory, we obtain
\begin{equation}
| f(\phi) | = f_0 e^{-\mu \phi_0} =  \Lambda^{-4} \gsim  3.9 \times 10^{15} \, {\rm eV^{-4}}\,,
 \label{DBIonic}
\end{equation}
for $g=0.1$.
Since we assume that  the scalar field is the source of dark energy, we get
\begin{equation*}
\rho_{\rm DE} = \rho_{\phi} =\frac{\gamma_0^2}{\gamma_0 + 1} \dot\phi |_0^2 + V(\phi_0) \simeq V(\phi_0) \,,
\end{equation*}
where we have used the fact that
 the kinetic term is very small because $\gamma_0\ll 1$. 
As for the potential of the scalar field, from observations we have constraint  as
\begin{equation}
V(\phi) = V_0 e^{-\lambda \phi_0} \approx 2.6 \times 10^{-47} \, {\rm GeV^4} \,. 
\label{DE}
\end{equation}
The multiplication of Eqs.~(\ref{DBIonic}) and (\ref{DE}) gives
\begin{equation*}
f_0 V_0 \approx  1.0 \times 10^{5} \,.
\end{equation*}
Surprisingly, this is the same order of magnitude in order to satisfy the coincidence problem discussed above. 

In Fig. \ref{F4}, we present the parameter space ($\lambda, g$) where we may 
find the solution for the dark energy problem.
\begin{figure}[h]
	\includegraphics[width=6cm]{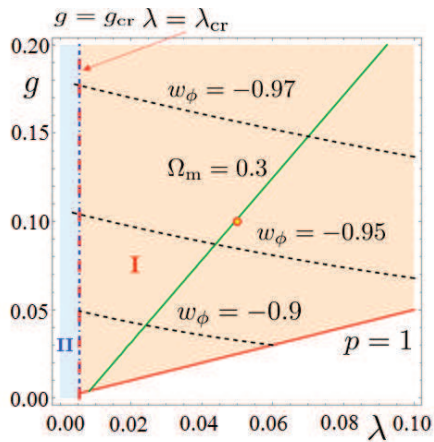} 
\caption{The existence regions of two accelerating 
solutions {\bf I} and {\bf II}, in the parameter space ($\lambda, g$) for the D-BIonic ($\epsilon=-1$).
We set $f_0V_0=10^5$.
The red dashed curve ($g=g_{\rm cr}$) and 
the blue dot-dashed line ($\lambda=\lambda_{\rm cr}$) are almost the same. 
The green curve denotes $\Omega_{\rm m}=0.3$, while the black dashed lines give $w_\phi=-0.97, -0.95$ and $-0.9$, respectively  from the above. The red solid lines denote $p=1$.  The red circle corresponds to the parameters we adopted ($\lambda=0.05, g=0.1$).} 
\label{F4}
\end{figure}

Using $f_0 V_0 = 10^5$, $\lambda = 0.05$, and $g = 0.1$, which is shown
by the red circle  in Fig. \ref{F4}, 
we obtain
\begin{align*}
\Omega_{\phi} &\approx 0.697 \,, \\
\Omega_{\rm m} &\approx 0.303 \,, \\
w_{\phi} &\approx - 0.96 \,.
\end{align*}
These are dark energy density, dark matter density, and the EOS parameter of dark energy of the Universe today. 
Note that the  eigenvalues of this fixed point is
%are
\begin{equation*}
\omega_\pm\approx -1.25 \pm 3.415 \times 10^4 \;i\,,
\end{equation*}
and then this solution is of course stable. 
This also shows the typical time scale to approach the solution {\bf I} is one e-folding time.
This means that by the D-BIonic gravity theory we may be able to solve the coincidence problem,
 which is difficult to realise in the original coupled quintessence model.

%======================================%
%<<<<<<<<<<<< SECTION VI  >>>>>>>>>>>>>>%
%======================================%
%%%%%%%%%%%%%%%%%%%%%%%%%%%%%%%%%%%%%%%%%%%%%%%%%%%
%%%%%%%%%%%%%%%%%%%%%%%%%%%%%%%%%%%%%%%%%%%%%%%%%%%
%%%%%%%%%%%%%%%%%%%%%%%%%%%%%%%%%%%%%%%%%%%%%%%%%%%
\section{Conclusions} 
\label{conclusions}
%%%%%%%%%%%%%%%%%%%%%%%%%%%%%%%%%%%%%%%%%%%%%%%%%%%
%%%%%%%%%%%%%%%%%%%%%%%%%%%%%%%%%%%%%%%%%%%%%%%%%%%

In this work, we study the cosmological dynamics of the D-BIonic and DBI scalar field coupled with matter fluid.
We assume the exponential forms for the potential and the coupling functions.
We find two interesting analytic solutions of the D-BIonic 
%gravity 
theory as well as the DBI theory,
which describe the accelerated expansion of the Universe.
One is similar to the conventional quintessence in the DBI theory, and the scalar field energy density becomes dominant.
The other one is a new scaling solution 
because it is found from non-canonical kinetic term as well as the matter coupling term.
This gives non-zero density parameter $\Omega_{\rm m}$, whose value depends on the coupling constants.
For the original coupled quintessence model, although we have a solution that may solve the coincidence problem, 
there is some difficulty such that it requires a large coupling constant between dark energy (scalar field) and matter fluid,
which is inconsistent with observational data from CMB. 
However, in the case of the D-BIonic 
%gravity 
theory, 
we find a successful coupled quintessence model by use of a newly found scaling solution with small coupling constant $g$. 
This may naturally solve the coincidence problem as well because the density parameter $\Omega_{\rm m}$ is 
the value of the attractor solution.
The solution is expected to  satisfy 
the observational data of the Universe for dark energy as well as the solar system constraint for the screening.
We find that the D-BIonic 
%gravity 
can solve both the dark energy problem and the coincidence problem.

Finally, since our analysis is only the background behaviour of the Universe, we have to analyse the details  furthermore, including 
the evolution of density perturbations as well as the CMB data, in order to confirm our model.
%Finally, 
Furthermore this work has been based on exponential forms of the inverse D3-brane-like tension $f(\phi)$, 
the potential term $V(\phi)$, and the conformal factor $A(\phi)$,
which can be extended
%We may extend them 
to be power-law functions. 
We leave 
%it 
them for future works.

%%%%%%%%%%%%%%%%%%%%%%%%%%%%%%%%%%%%%%%%%%%%%%%%%%%
%%%%%%%%%%%%%%%%%%%%%%%%%%%%%%%%%%%%%%%%%%%%%%%%%%%
\section*{Acknowledgement}
%%%%%%%%%%%%%%%%%%%%%%%%%%%%%%%%%%%%%%%%%%%%%%%%%%%
%%%%%%%%%%%%%%%%%%%%%%%%%%%%%%%%%%%%%%%%%%%%%%%%%%%
S.P. is supported by Japanese Government (Monbukagakusho) Scholarship.
This work was supported in part by Grants-in-Aid from the 
Scientific Research Fund of the Japan Society for the Promotion of Science 
No. 16K05362 (K.M.) and  No. 16K17709 (S.M.).

\newpage

%===============================%
%<<<<<<<<<<<< APPENDIX >>>>>>>>>>>>>>%
%======================================%
%%%%%%%%%%%%%%%%%%%%%%%%%%%%%%%%%%%%%%%%%%%
%%%%%%%%%%%%%%%%%%%%%%%%%%%%%%%%%%%%%%%%%%%
%%%%%%%%%%%%%%%%%%%%%%%%%%%%%%%%%%%%%%%%%%%
%%%%%%%%%%%%%%%%%%%%%%%%%%%%%%%%%%%%%%%%%%%
\appendix
%%%%%%%%%%%%%%%%%%%%%%%%%%%%%%%%%%%%%%%%%%%%%%%%%%%
%%%%%%%%%%%%%%%%%%%%%%%%%%%%%%%%%%%%%%%%%%%%%%%%%%%
%%%%%%%%%%%%%%%%%%%%%%%%%%%%%%%%%%%%%%%%%%%%%%%%%%%

%%%%%%%%%%%%%%%%%%%%%%%%%%%%%%%%%%%%%%%%%%%
%%%%%%%%%%%%%%%%%%%%%%%%%%%%%%%%%%%%%%%%%%%
\section{D-BIonic screening mechanism}
%%%%%%%%%%%%%%%%%%%%%%%%%%%%%%%%%%%%%%%%%%%
%%%%%%%%%%%%%%%%%%%%%%%%%%%%%%%%%%%%%%%%%%%

The D-BIonic screening mechanism is 
obtained from the Dirac-Born-Infeld (DBI)-like action:
\begin{align}
S_{\phi} = \int d^4 x \sqrt{-g} \left[\Lambda^4 \left(\sqrt{1- \Lambda^{-4} (\partial \phi)^2} -1\right)\right] \,,
\label{action0}
\end{align}
where $\Lambda$ is a characteristic mass scale.
The overall sign in above action has been flipped from the original DBI action, which is necessary to obtain a screening mechanism. 
This theory does not contain ghosts because the kinetic term still has a correct sign. 
We assume that the scalar field couples conformally to matter fluid.  The equation of motion of the scalar field is given by
\begin{equation}
\nabla_{\mu} \left(\frac{\nabla^{\mu}\phi}{\sqrt{1 -\Lambda^{-4} (\partial\phi)^2}}\right)= - \frac{g}{M_{\rm PL}} T^{\rm (m)} \,, 
\label{eom0}
\end{equation}
where $g$ is a dimensionless coupling constant, $M_{\rm PL}$ is the reduced Planck mass, 
and $T^{\rm (m)}$ is the trace of matter energy 
momentum tensor.
The above equation can be divided into a linear term and a non-linear term, which leads to a screening 
mechanism as we mentioned. 
For a static point source at the origin, $T^{\rm (m)} = - M \delta (\vec x)$
($M$: the mass of the point source),  under the static and spherically
 symmetric assumptions, we find a solution as
\begin{equation*}
\frac{d\phi}{dr} = \frac{\Lambda^2}{\sqrt{1 + (r/r_{\ast})^4}} \,,
\end{equation*}
where 
\begin{equation*}
r_{\ast} \equiv \frac{1}{\Lambda} \left(\frac{g M}{4\pi M_{\rm PL}}\right)^{1/2} \,,
\end{equation*}
which gives a typical length scale just like 
%as
the Vainshtein radius. 
This radius is proportional to $M^{1/2}$, whereas the Vainshtein radius is proportional 
to $M^{1/3}$. Deep inside this radius $r \ll r_{\ast}$, the solution becomes
\begin{equation*}
\frac{d \phi}{dr} \simeq \Lambda^2 \,,
\end{equation*}
then the fifth force comparing to the Newtonian force is
\begin{equation*}
F_{\phi}/F_N \simeq 2g^2 \left(\frac{r}{r_{\ast}}\right)^2 \ll 1\,,
\end{equation*}
where 
\begin{eqnarray*}
F_\phi&=&-{g\over M_{\rm PL}}{d\phi\over dr}\,,
\\
F_N&=&-{M\over 8\pi M_{\rm PL}^2\, r^2}
\,.
\end{eqnarray*}
Thus, the force is screened and Newtonian gravity is recovered. 
On the other hand, far from the source $r \gg r_{\ast}$, we find the solution as
\begin{equation*}
\frac{d\phi}{dr} \simeq \Lambda^2 \left(\frac{r_{\ast}}{r}\right)^2 \,,
\end{equation*}
which gives
\begin{equation*}
F_{\phi}/F_N \simeq 2g^2 \,.
\end{equation*}
Therefore, the fifth force is unscreened at large distance.

%%%%%%%%%%%%%%%%%%%%%%%%%%%%%%%%%%%%%%%%%%%
%%%%%%%%%%%%%%%%%%%%%%%%%%%%%%%%%%%%%%%%%%%
\section{Other Analytic Solutions }
\label{other_solutions}
%%%%%%%%%%%%%%%%%%%%%%%%%%%%%%%%%%%%%%%%%%%
%%%%%%%%%%%%%%%%%%%%%%%%%%%%%%%%%%%%%%%%%%%

In this appendix, we present the other analytic solutions for the case {\bf III} and {\bf IV}.

%%%%%%%%%%%%%%%%%%%%%%%%%%%%%%%%%%%%%%%%%%%%%%%%%%%%%%%%
%%%%%%%%%%%%%%%%%%%%%%%%%%%%%%%%%%%%%%%%%%%%%%%%%%%%%%%%
\subsection{Case III :  $\lambda/\mu<-1$ and $q=2$}
%%%%%%%%%%%%%%%%%%%%%%%%%%%%%%%%%%%%%%%%%%%%%%%%%%%%%%%%
%%%%%%%%%%%%%%%%%%%%%%%%%%%%%%%%%%%%%%%%%%%%%%%%%%%%%%%%
In this case, we find from Eqs. (\ref{Veom}) and (\ref{rhoeom}) 
\beann
&&
-2\mu\gamma_0[3p(\gamma_0+1)-2\gamma_0]
%\\
%&&
%~~
+g[3p^2(\gamma_0+1)\mu^2-4\gamma_0^2]=0 \,,
\\
&&
\rho_0 t_0^2
={1\over \mu^2(\gamma_0+1)[\lambda+g]}
\\
&&
~~\times
\{2\mu\gamma_0[3p(\gamma_0+1)-2\gamma_0]
%\\
%&&
%~~~~~~~~
+\lambda(3p^2(\gamma_0+1)\mu^2-4\gamma_0^2)
\}
\,,
\enann
and we obtain from the definition of $q$ 
\begin{equation*}
p={2\over 3}\left(1-{g\over \mu}\right)
\,.
\end{equation*}
These equations provide the value of  $\gamma_0$ as
\begin{align*}
&
\gamma_0={g(\mu-g)\over 3-g(\mu-g)}
\,,
\end{align*}
which turns out that $\gamma_0$ is negative because $\mu<0$ (we choose $\lambda > 0$). 
Then we do not have a solution in this case unless $V_0=0$.
If there is no potential, $\mu$ can be positive, and then the above solution solves our basic equations.
Actually, in this case, we can find an exact solution where the matter and kinetic term of the 
DBI field scales, which is the extension of the solution found in \cite{Copeland} to include the effect 
of matter coupling. Regardless of this, since this does not give an accelerating Universe, we do not consider
this solution any more in this paper.

%%%%%%%%%%%%%%%%%%%%%%%%%%%%%%%%%%%%%%%%%%%%%%%%%%%%%%%%
%%%%%%%%%%%%%%%%%%%%%%%%%%%%%%%%%%%%%%%%%%%%%%%%%%%%%%%%
\subsection{Case IV : $\lambda/\mu<-1$ and $q>2$}
%%%%%%%%%%%%%%%%%%%%%%%%%%%%%%%%%%%%%%%%%%%%%%%%%%%%%%%%
%%%%%%%%%%%%%%%%%%%%%%%%%%%%%%%%%%%%%%%%%%%%%%%%%%%%%%%%

From Eq. (\ref{Veom}) and (\ref{rhoeom}), at late time we find two equations:
\begin{align*}
&-2\mu\gamma_0[3p(\gamma_0+1)-2\gamma_0] +g [3p^2(\gamma_0+1)\mu^2-4\gamma_0^2] \\
&~~~=0 \,,
\\
&2\mu\gamma_0[3p(\gamma_0+1)-2\gamma_0]+\lambda[3p^2(\gamma_0+1)\mu^2-4\gamma_0^2]=0
\,.
\end{align*}
Then, we find
\begin{align*}
p^2&={4\gamma_0^2\over 3(\gamma_0+1)\mu^2 } \,,
\\
p&={2\gamma_0\over 3(\gamma_0+1)}
\,.
\end{align*}
These equations give 
\begin{align*}
p&={2\over 3 } \left(1- {3\over \mu^2}\right) \,, \\
\gamma_0&={\mu^2 -3\over 3}
\,,
\end{align*}
from which we obtain
\begin{equation*}
\phi_0 = -{1\over \mu}\ln {\mu^4(\mu^2-6) t_0^2 \over 4\epsilon f_0(\mu^2-3)^2}
\,.
\end{equation*}
Since $\gamma_0>0$, $\mu^2$ must be larger than 3. We also find 
\\
(i) $\mu^2>6$, and then $1/3<p<2/3$ for the DBI theory, whereas \\
(ii) $3<\mu^2<6$, and then $0<p<1/3 $ for D-BIonic theory.

In both cases, since $p<2/3$, we cannot find the accelerating 
expansion. Note that the EOS parameter of the scalar field is
\begin{equation*}
w_\phi\equiv\frac{P_\phi}{\rho_\phi}={1\over \gamma_0}
\,,
\end{equation*}
which is positive definite. 
For the D-BIonic theory, $w_\phi>1$, which gives a ``supersonic'' flow.
It is why we find $p<1/3$.

%%%%%%%%%%%%%%%%%%%%%%%%%%%%%%%%%%%%%%%%%%%%%%%%%%
%======================================%
%<<<<<<<<<<<< BIBLIOGRAPHY >>>>>>>>>>>>%
%======================================%
%%%%%%%%%%%%%%%%%%%%%%%%%%%%%%%%%%%%%%%%%%%%%%%%%%
%%%%%%%%%%%%%%%%%%%%%%%%%%%%%%%%%%%%%%%%%%%%%%%%%%

\end{document}